\documentclass[5p,twocolumn,num]{elsarticle}

\makeatletter
\def\ps@pprintTitle{%
	\let\@oddhead\@empty
	\let\@evenhead\@empty
	\def\@oddfoot{\centerline{\thepage}}%
	\let\@evenfoot\@oddfoot}
\makeatother
\usepackage{amssymb}

\usepackage{algpseudocode}
\usepackage{algorithm}
\newfloat{alg}{thp}{lop}
\floatname{algorithm}{Algorithm}
\usepackage{natbib}
\usepackage{amsmath, amsfonts, mathrsfs,mathtools}
\usepackage{graphicx}
\usepackage{array}
\usepackage{color}
\usepackage{soul}
\usepackage{caption}
\usepackage{epstopdf}
\usepackage{mathtools}
\usepackage{algpseudocode}
\usepackage{algorithm}
\usepackage{subcaption}
\usepackage{float}
\usepackage[caption = false]{subfig}
\usepackage{adjustbox}
\usepackage{multirow}
\usepackage{enumitem}
\usepackage{multirow}
\usepackage{booktabs}
\usepackage{lipsum} 

\setlist{nolistsep}

\newtheorem{remark}{Remark}
\newtheorem{example}{Example}

\newtheorem{prob}{Problem}
\newtheorem{theorem}{Theorem}
\newtheorem{proof}{Proof}

\begin{document}
	\begin{frontmatter}
		\sffamily	

\title{A Data-Ensemble-Based Approach for Sample-Efficient LQ Control of Linear Time-Varying Systems}

\author[1]{Sahel Vahedi Noori}
\author[2]{Maryam Babazadeh\corref{cor1}}

\address[1]{University of Michigan, Ann Arbor, USA}
\address[2]{Sharif University of Technology, Tehran, Iran}

\cortext[cor1]{Corresponding author: babazadeh@sharif.edu}
		
\begin{abstract}
	\sffamily
This paper presents a sample-efficient, data-driven control framework for finite-horizon linear quadratic (LQ) control of linear time-varying (LTV) systems. In contrast to the time-invariant case, the time-varying LQ problem involves a differential Riccati equation (DRE) with time-dependent parameters and terminal boundary constraints, complicating data-driven control. Additionally, the time-varying dynamics invalidate the use of the Fundamental Lemma. To overcome these challenges, we formulate the LQ problem as a nonconvex optimization problem and conduct a rigorous analysis of its dual structure. By exploiting the inherent convexity of the dual problem and analyzing the KKT conditions, we derive an explicit relationship between the optimal dual solution and the parameters of the associated Q-function in time-varying case. This theoretical insight supports the development of a novel, sample-efficient, non-iterative semidefinite programming (SDP) algorithm that directly computes the optimal sequence of feedback gains from an ensemble of input-state data sequences without requiring model identification or a stabilizing controller. The resulting convex, data-dependent framework provides global optimality guarantees for completely unknown LTV systems. As a byproduct, the finite-horizon LQ control of linear time-invariant (LTI) systems emerges as a special case of the proposed method. In this setting, a single input-state trajectory suffices to identify the optimal LQ feedback policy, improving significantly over existing Q-learning approaches for finite horizon LTI systems that typically require data from multiple episodes. The approach provides a new optimization-based perspective on Q-learning in time-varying settings and contributes to the broader understanding of data-driven control in non-stationary environments. Simulation results show that, compared to recent methods, the proposed approach achieves superior optimality and sample efficiency on LTV systems, and indicates potential for stabilizing and optimal control of nonlinear systems.
\end{abstract}

	\end{frontmatter}

\section{Introduction} \label{sec1}

Optimal control of unknown dynamic systems is typically approached from two perspectives: model-based and model-free control. Model-based methods rely on fitting a system model using historical data, where the model is used for control design but the data itself is not directly embedded in the control law. In contrast, model-free methods avoid explicit model identification and directly learn the optimal control policy. Interest in model-free control is growing due to the increasing complexity and inaccessibility of real-world dynamics. Moreover, the classical two-step model-based procedure is often suboptimal, as system identification and control design lack a separation principle in general. As a result, integrated model-free approaches can potentially outperform model-based methods \cite{markovsky2023data}, especially when data is insufficient for accurate modeling \cite{van2020data}.

Recently, there has been a surge of interest in applying model-free control paradigm to tackle optimal control challenges, particularly through the lens of reinforcement learning algorithms. For example, in \cite{kiumarsi2014reinforcement} and \cite{li2018off}, Q-learning algorithms are introduced for optimal control of discrete-time and continuous-time LTI systems, respectively. In \cite{lee2018primal}, a model-free primal-dual Q-learning algorithm is proposed for LQR control of LTI systems. The method in \cite{lee2018primal} assumes the availability of state trajectories collected under various control policies with the same initial conditions. An alternative model-based primal-dual algorithm for stochastic LQR control of LTI systems, along with its model-free implementation, is presented in \cite{li2022model}. This approach is closely aligned with the Policy Iteration (PI) algorithm, eliminating the need for an excitation signal. Additionally, \cite{fazel2018global,hu2022towards} introduce non-convex model-free policy gradient algorithms for LQR control, demonstrating its global convergence to the optimal solution and sample efficiency.\\
Recent studies highlight that stabilization and certain classes of optimal control problems for LTI systems can be effectively addressed using model-free linear matrix inequalities (LMIs) \cite{de2019formulas, van2021matrix}. In the context of model-free LMIs, a convex and non-iterative  LQR design is proposed in \cite{farjadnasab2022model}. This method successfully derives the exact $Q$ function for the LQR problem with an infinite-horizon cost function in LTI systems. Alternatively, \cite{de2021low} proposes sufficient conditions, yielding a data-dependent LMI-based stabilizing controller with a guaranteed relative error.
In \cite{calafiore2020output}, the finite horizon LQ problem for LTI systems is reformulated as a sequence of backward data-dependent SDPs. It is known that if system parameters change over time, sequential backward methods may suffer from mismatch between stages.\\
Although most existing model-free control studies based on optimization focus on LTI systems, LTV systems are better suited for capturing time-dependent behaviors seen in many applications including aerospace, energy systems, robotics, and chemical processes. They are also widely used to approximate nonlinear dynamics through local linearization.
Some contributions explore the integration of model-free optimization techniques, such as extremum seeking \cite{Liu2017BatchtoBatchFL, scheinker2021extremum}, in time-varying systems. For the special case of discrete periodic LTV dynamics, \cite{yang2018efficient} reformulates the LQR control as an optimal control problem for an augmented LTI model through a lifting procedure. Additionally, a learning-based adaptive algorithm is introduced in \cite{pang2020reinforcement} to design stabilizing approximate optimal controllers for continuous-time linear periodic systems. Meanwhile, in \cite{qu2021stable}, an online iterative linear quadratic control is proposed to ensure stability for a specific class of LTV systems while simultaneously minimizing control costs in a sub-optimal manner.\\
Approximate solutions for general discrete-time LTV systems are presented in \cite{pang2018data, pang2019adaptive}, employing Policy Iteration and Value Iteration algorithms. In \cite{possieri2021iterative}, an approximate dynamic programming solution to LQ optimal problems for time-varying dynamics in terms of Q-factors is proposed.
For the case of continuous-time LTV systems, \cite{fong2018dual} introduces a dual loop iterative algorithm that converges to an arbitrarily small neighborhood of the optimal solution provided that the discretization parameter and the excitation signals are judiciously chosen. Recently, a data-dependent set of LMIs has been introduced in \cite{nortmann2020data,nortmann2021direct} to design state feedback controllers for LTV systems based on covariance selection problem. The results are an extension of data-driven covariance selection problem for LTI systems \cite{rotulo2020data} to LTV systems.\\

Solving the model-free finite-horizon LQ control problem for nonstationary linear systems involves unique challenges that set it apart from the classical LTI case. The time-varying nature of system matrices invalidates powerful tools such as Willems’ Fundamental Lemma~\cite{willems2005note}, which underpins many data-driven methods for LTI systems. Although limited extensions exist for specific classes like periodic and affine parameter-varying systems~\cite{verhoek2021fundamental}, no generalization applies to arbitrary LTV dynamics. Furthermore, standard model-free and RL approaches typically assume a stationary environment, limiting their ability to ensure optimality or stability in time-varying settings. Existing value-based methods for LTV systems often exhibit sample inefficiency and sub-optimal performance \cite{pang2018data, pang2019adaptive,possieri2021iterative}. As a result, much of the current research focuses on specific subclasses of time-varying systems~\cite{yang2018efficient,pang2020reinforcement,qu2021stable}. These limitations underscore the need for new, sample-efficient, and identification-free control frameworks specifically designed for general LTV systems.

This work introduces a novel, sample-efficient, and convex approach to finite-horizon LQ control for LTV systems, with the following key contributions: (i) A novel single-step SDP-based formulation for the finite-horizon LQ control of LTV systems is proposed. To the best of our knowledge, this reformulation is entirely new. (ii)
Optimality is established through a novel primal-dual interpretation, and it is shown that the optimal solution in the dual problem corresponds to the optimal time-varying $Q$-function in Q-learning framework. The method provides a direct convex characterization of the Q-function for finite-horizon LTV systems, laying the groundwork for principled Q-learning approaches in this underexplored setting. (iii) The method only requires $n+m$ batches of input-output data, avoids iterative policy updates or backward optimizations, making it well-suited for implementation in reinforcement learning setups. (iv) As a byproduct, the finite-horizon LQ control of LTI systems is treated as a special case of the presented method. It is shown that, in this setting, a single input-state trajectory of length $n+m+1$ is sufficient to identify the optimal LQ feedback policy, significantly improving over existing Q-learning approaches based on sequential backward solving that require multiple episodes \cite{calafiore2020output, possieri2021iterative}.

The remainder of this paper is structured as follows. In Section \ref{sec2}, an equivalent non-convex optimization problem for the finite-horizon LQ problem in LTV systems is derived. Section \ref{sec3} employs Duality Theory and KKT conditions to formulate a model-based semi-definite program for the optimal control of LTV systems. Subsequently, in Section \ref{sec4}, a model-free semi-definite program is introduced by incorporating an ensemble of input-state trajectories into the developed SDPs. This model-free approach is guaranteed to yield a globally optimal solution for the sequence of control gains. To showcase the effectiveness and accuracy of the proposed method, we present simulation results in Section \ref{sec5}.\\

\textbf{Notation:}  The symbol $\mathbb{Z}$ denotes the sets of non-negative integers. The notation $\mathbb{R}^n$ represents the $n$-dimensional Euclidean space, and $\mathbb{R}^{n\times m}$ refers to the set of all $n\times m$ real matrices. 
Furthermore, $\mathbb{S}^n$,  $\mathbb{S}_+^n$, and  $\mathbb{S}_{++}^n$ denote symmetric $n\times n$ matrices, cone of symmetric $n \times n$ positive semi-definite matrices, and symmetric $n \times n$ positive definite matrices, respectively. The symbol $I_n$ represents the $n\times n$ identity matrix. In addition, the expressions $X \succ 0, X \prec 0, X\succeq 0,$ and $X\preceq 0$ denote symmetric positive definite, symmetric negative definite, symmetric positive semi-definite, and symmetric negative semi-definite matrices, respectively.\\
For a signal $u(k):~\mathbb{Z} \rightarrow \mathbb{R}^m$, the sequence \( \{u(i),...,u(j)\}\) with $i \leq j$ is given by $u_{[i,j]}$. Similarly, for the time-varying matrix $K(k):~\mathbb{Z} \rightarrow \mathbb{R}^{m \times n}$, the sequence \( \{K(i),...,K(j)\}\) with $i \leq j$ is refered to as $K_{[i,j]}$.

\section{Problem setup} \label{sec2}
Consider the discrete-time LTV system,
\begin{align}
\nonumber	&  x(k+1)=A(k)x(k)+B(k)u(k),\\
& x(0) :=z\in \mathbb{R}^n,~~~
    k\in\mathbb{Z}
    \label{eq:1}
\end{align}
where \(x(k)\in\mathbb{R}^n\) is the state vector, \(u(k)\in\mathbb{R}^m\) is the input vector, and \(A(k)\in\mathbb{R}^{n\times n}\) and \(B(k)\in\mathbb{R}^{n\times m}\) characterize the time-varying dynamics of the system. The objective of Finite-horizon discrete-time LQ problem for LTV system \eqref{eq:1} is to determine the optimal control sequence \(u^*_{[0,N-1]}= \{u^*(0), u^*(1), ..., u^*(N-1)\}\) such that the finite-cost function \eqref{eq:2} is minimized.
\begin{align}\label{eq:2}
 \nonumber   & J\big(u_{[0,N-1]},x(0)\big) = \gamma^{N}x^{T}(N)Q_{f}x(N)\\
    &+ \sum_{k=0}^{N-1} \gamma^{k}\big[x^{T}(k)Q(k)x(k)
    +u^{T}(k)R(k)u(k)\big].
\end{align}
In  \eqref{eq:2}, \(Q_f\in\mathbb{S}^n_{+}\), \(Q(k)\in\mathbb{S}^n_{+}\) and \(R(k)\in\mathbb{S}^m_{++}\) for {$k=0,1,...,N-1$}, are the weighting matrices, and $\gamma \in (0,1]$ is a discount factor that determines the relative weighting of future stage costs in the objective function. A value of $\gamma = 1$ corresponds to the undiscounted case, while smaller values assign less importance to later time steps. \\
The optimal policy with the cost function \eqref{eq:2} can be obtained by using dynamic programming \cite{bertsekas2015dynamic} as the time-varying state-feedback control,
\begin{equation} \label{eq:3}
    u(k)=K(k)x(k),
\end{equation}
with \(K(k)\in\mathbb{R}^{m\times n}\) for \(k=0, 1,..., N-1\).  Let $z_i$ for \(i=\{1,2,...,r\}\) constitutes a collection of $r \geq n$ linearly independent vectors. These vectors serve as distinct initial points for state trajectories.  By integrating \eqref{eq:3} into \eqref{eq:2}, the canonical LQ control problem can be redefined as finding the optimal feedback gain sequence \( K^*_{[0,N-1]}=\{K^*(0), K^*(1), ..., K^*(N-1)\}\)  through the solution of the optimization problem,
\begin{equation}
    \textrm{minimize} \;\; \sum_{i=1}^{r} J\big(K_{[0,N-1]},z_i\big).
    \label{eq:100}
\end{equation}
It is known that distinct initial conditions yield the same optimal solution \(K^*_{[0,N-1]}= \{K^*(0), K^*(1), ..., K^*(N-1)\}\) \cite{kirk2004optimal}, enabling the representation of problem \eqref{eq:2} with arbitrary initial conditions. However, in \eqref{eq:100} the set of initial states $\{z_i\}$  spans the entire state space. This coverage ensures that the cost function in \eqref{eq:100} reflects the system’s behavior across all directions of the initial state, thereby making the optimization problem representative and well-posed for the primal-dual analysis in Section~\ref{sec3}.

Given a fully known dynamics of the LTV system \eqref{eq:1}, the optimal time-varying sequence of controllers $K^*(k)$ is derived as \cite{bertsekas2015dynamic},
\begin{equation}
    \begin{split}
    K^*(k)&=-\big(R(k)+\gamma B^T(  k)P(k+1)B(k)\big)^{-1} \\
    & \times\gamma B^T(k)P(k+1)A(k),
    \label{eq:101}
    \end{split}
\end{equation}
where \( P(k) \) for \(k=0,1,...,N-1\), is the solution of the Difference Riccati Equations,
    \begin{align}\label{eq:73}
 \nonumber   P(k) &= Q(k)+\gamma A^T(k)P(k+1  )A(k)-\gamma^2 A^T(k)\\
  \nonumber  &\times P(k+1)B(k)\big(R(k)+\gamma  B^T(k)P(k+1)B(k)\big)^{-1}\\
    & \times B^T(k)P(k+1)A(k),
    \end{align}
with \( P(N)=Q_{f} \). 
Development of the control law begins at the terminal time $N$ and proceeds backward, by recursive solving of \eqref{eq:73}. The seminal results of \eqref{eq:101} and \eqref{eq:73} are widely used in model-based optimal control, i.e., when the system matrices $A(k)$ and $B(k)$ are known for \(k=0,1,...,N-1\). The linearity of the optimal control law in this problem is preserved
even for the cases where the state $x(k)$ is not completely observable \cite{bertsekas2015dynamic}.

To obtain the equivalent optimal sequence of solutions independent of the dynamical model \eqref{eq:1}, define the augmented vector $v\big(k;K(k),z_i\big)$ as follows,
\begin{equation}
    v\big(k;K(k),z_i\big)=\gamma^{\frac{k}{2}}
    \begin{bmatrix}
    x(k)\\
    u(k)
    \end{bmatrix}=\gamma^{\frac{k}{2}}
    \begin{bmatrix}
    x(k)\\
    K(k)x(k)
    \end{bmatrix}\in\mathbb{R}^{n+m}
    \label{eq:4}
\end{equation}
for \(k=0,1,...,N-1\). Based on \eqref{eq:4}, the objective function \eqref{eq:2} can be expressed as,
\begin{align}\label{eq:J1}
 \nonumber  & J\big(K_{[0,N-1]},z_i\big)=  \gamma^{N}x^{T}(N)Q_{f}x(N)\\
 \nonumber  & +\sum_{k=0}^{N-1}v^{T}\big(k;K(k),z_i  \big)\Lambda(k) v\big(k;K(k),z_i\big)\\
  \nonumber &  =v^{T}\big(N-1;K(N-1),z_i\big)E^{T}(N  -1)Q_{f}E(N-1)\\
    \nonumber  & \times v\big(N  -1;K(N-1),z_i\big) \\
   & +\sum_{k=0}^{N-1}v^{T}\big(k;K(k),z_i  \big)\Lambda(k) v\big(k;K(k),z_i\big),
\end{align}
with,
\begin{equation}
    \Lambda(k)= \begin{bmatrix}
    Q(k) \;& 0\\
    0 \;& R(k)
    \end{bmatrix}, \qquad  E(k)=\sqrt\gamma\begin{bmatrix}
    A(k) \;\;& B(k) 
    \end{bmatrix}.
    \label{eq:7}
\end{equation}
Accordingly, the minimization problem \eqref{eq:100} can be expressed as, 
\begin{equation}\label{eq:J2}
    \begin{split}
    \text{minimize~~} &\sum_{i=1}^{r} v^{T}\big(N-1  ;K(N-1),z_i\big)E^{T}(N-1)Q_{f}\\
   & \times E(N  -1)v\big(N-1;K(N-1),z_i\big)\\
   & +\sum_{i=1}^{r}\sum_{k=0}^{N-1}v^{T}\big(k  ;K(k),z_i\big)\Lambda(k) v\big(k;K(k),z_i\big).       
    \end{split}
\end{equation}
Furthermore, deriving the dynamics in terms of the augmented signal $v\big(k;K(k),z_i\big)$ is a straightforward process, yielding,
\begin{equation}    v\big(k+1;K(k+1),z_i\big)=A_K(k)v\big(k;K(k),z_i\big),
    \label{eq:9}
\end{equation}
for \(k=0,1,...,N-2\), with,
\begin{equation}
    A_K(k)=\sqrt{\gamma}
    \begin{bmatrix}
    A(k) \;\;& B(k)\\
    K(k+1)A(k) \;\;& K(k+1)B(k)
    \end{bmatrix}.
    \label{eq:10}
\end{equation}
Next, define the positive semi-definite matrix $S(k)$ as,
\begin{equation}
    S(k)=\sum_{i=1}^{r}v\big(k;K(k),z_i\big) v^{T}\big(k;K(k),z_i\big)\succeq 0,
    \label{eq:13}
\end{equation}
Then, for $k=0,1,...,N-2$,
\begin{equation}
    S(k+1)=A_K(k)S(k)A_K^T(k), 
    \label{eq:14}
\end{equation}
\begin{align} \label{eq:15_new}
  \nonumber   S(0) & = \sum_{i=1}^{r}\begin{bmatrix}
    z_i\\
    K(0)z_i
    \end{bmatrix} 
    \begin{bmatrix}
    z_i\\
    K(0)z_i
    \end{bmatrix}^{T}\\
   & =\begin{bmatrix}
    I_{n}\\
    K(0)
    \end{bmatrix}Z
    \begin{bmatrix}
    I_{n}\\
    K(0)
    \end{bmatrix}^{T},
\end{align}
with aggregated initial state matrix $Z=\sum_{i=1}^{r}z_iz^{T}_i$. From a covariance perspective, the matrix $Z$ can be interpreted as a scaled unnormalized empirical covariance matrix of the initial states, under the assumption that the mean is zero. Given that the vectors $z_i$ are linearly independent and the condition $r \geq n$ holds, i.e., the set of initial states spans the entire state space, the matrix $Z$ is of full rank and belongs to the positive definite cone $\mathbb{S}_{++}^n$.

The minimization problem \eqref{eq:J2} can be alternatively formulated as,
\begin{align} \label{eq:J3}
\nonumber    \text{minimize~~} & \text{trace}\big(E^{T}(N-1)Q_{f}E(N-1)S(N-1)\big)\\
    &+\sum_{k=0}^{N-1}\text{trace}\big(\Lambda(k)S(k)\big).
\end{align}
Putting together the non-convex minimization problem \eqref{eq:J3} and the constraints in \eqref{eq:14} and \eqref{eq:15_new}, the primary optimization problem can be expressed as Problem 1. In this context, the term ``equivalent'' signifies that if $K^*(k)$ with \(k=0,1,...,N-1\) denotes the optimal solution for problem \eqref{eq:J2}, then $(K^*(k),S^*(k))$ is guaranteed to be the optimal solution for Problem 1. The uniquely determined value of $S^*(k)$ is specified by,
\begin{align}
    \nonumber & S^*(0)  = \begin{bmatrix}
    I_{n}\\
    K^*(0)
    \end{bmatrix}Z
    \begin{bmatrix}
    I_{n}\\
    K^*(0)
    \end{bmatrix}^{T},\\
   &  S^*(k+1) =A_{K^*}(k)S^*(k)A_{K^*}^T(k).
\end{align}
\\
\begin{prob}\label{problem1}
    Non-convex primal problem with decision variables \(S(k)\in\mathbb{S}_{+}^{n+m} \) and \(K(k)\in\mathbb{R}^{m\times n}\), indexed over $k= 0, 1, ..., N-1$.
\end{prob}
    \begin{align*}
& \textrm{minimize~~}  \text{trace}\big(E^{T}(N-1)Q_{f}E(N-1)S(N-1)\big)\\
    & ~~~~~~~~+\sum_{k=0}^{N-1}\text{trace}(\Lambda(k)S(k))\\
    &   \textrm{subject  to}: 
    S(0)=\begin{bmatrix}
    I_{n}\\
    K(0)
    \end{bmatrix}Z
    \begin{bmatrix}
    I_{n}\\
    K(0)
    \end{bmatrix}^{T}, \\
    & S(k+1)=A_K(k)S(k)A_K^T(k),~~ S(k)\succeq0,\\
    & A_K(k)=\sqrt{\gamma}
    \begin{bmatrix}
    A(k) \;\;& B(k)\\
    K(k+1)A(k) \;\;& K(k+1)B(k)
    \end{bmatrix},\\
    & \forall k \in \{0, 1, \dots, N-2\}.
\end{align*}  

\section{SDP-based finite-horizon LQ problem}\label{sec3} 

In this section, duality theory is utilized to reformulate the inherently non-convex Problem~\ref{problem1} as a convex optimization problem. Despite its non-convex nature, we demonstrate that the optimal solution of the associated dual problem can be used to recover the globally optimal sequence of LQ state-feedback gains $K^*_{[0,N-1]}$ for LTV systems. Additionally, we show that the choice of the positive definite matrix $Z$ influences only the objective values of the primal and dual problems, while leaving the optimal control gain sequence unaffected. We establish that the solution of the dual problem corresponds to the state-action value function (Q-function) of the finite-horizon LQ problem in the time-varying setting. Accordingly, the globally optimal Q-function matrix for the time-varying cas are derived explicitly. Finally, the dual formulation is cast as a model-based semidefinite program (SDP) with a favorable structure, which facilitates the development and integration of model-free optimization approaches.

Consider the Lagrangian of Problem \ref{problem1} with equality constraint multipliers $G(k)$ and inequality constraint multipliers $H(k)$ for \(k=0,1,...,N-1\) given by,

\begin{equation*}
\begin{split}
    &\mathcal{L}(K,S,G,H)=\text{trace}\big(E^{T}(N-1)Q_{f}E(N-1)\\
    &\;\;\;\;\times S(N-1)\big)+\sum_{k=0}^{N-1}\text{trace}(\Lambda(k) S(k))\\
&\;\;\;\;+\text{trace}\Bigg(\bigg(\begin{bmatrix}
    I_{n}\\
    K(0)
    \end{bmatrix}Z
    \begin{bmatrix}
    I_{n}\\
    K(0)
    \end{bmatrix}^{T}-S(0)\bigg)G(0)\Bigg)\\
    &\;\;\;\;+\sum_{k=0}^{N-2}\text{trace}\Big(\big(A_K(k)S(k)A_K^T(k)-S(k+1)\big)G(k+1)\Big)\\
    &\;\;\;\;-\sum_{k=0}^{N-1}\text{trace}\big( H(k)S(k)\big)\\ 
    \end{split}
\end{equation*}
\begin{equation}
    \begin{split}
    &=\sum_{k=0}^{N-2}\text{trace}\Big(\big(\Lambda(k)-G(k)+A_K^T(k)G(k+1)A_K(k)\\
    &\;\;\;\;-H(k)\big)S(k)\Big)+\text{trace}\Big(\big(\Lambda(N-1)-G(N-1)\\
    &\;\;\;\;+E^{T}(N-1)Q_{f}E(N-1)-H(N-1)\big)S(N-1)\Big)\\
    &\;\;\;\;+\text{trace}\bigg(Z\begin{bmatrix}
    I_{n}\\
    K(0)
    \end{bmatrix}^{T}G(0)
    \begin{bmatrix}
    I_{n}\\
    K(0)
    \end{bmatrix}\bigg).
    \end{split}
    \label{eq:18}
\end{equation}

The Lagrangian dual function is,
\begin{equation}
    g(G,H)=\underset{K(k),S(k)}{\text{inf}} \; \mathcal{L}(K,S,G,H),
    \label{eq:19}
\end{equation}
where \( K(k)\) and \( S(k)\) for \(k=0,1,...,N-1\), are the variables and the dual problem is,
\begin{equation}
    \underset{\underset{j=0,1,...,N-1}{H(j) \succeq 0,~ G(j)}}{\text{maximize}} \; g(G,H).
    \label{eq:20}
\end{equation}

\begin{theorem}\label{theorem1} The dual problem \eqref{eq:20} can be expressed as a convex maximization problem with conic constraints shown in Problem \ref{problem2}.
\end{theorem}
\begin{prob}\label{problem2}
    Dual problem with variables \( W\in\mathbb{S}^{n\times n} \) and \( G(k)\in\mathbb{S}^{n+m} \) indexed over \(k=0,1,...,N-1\).
    \begin{subequations}\label{problem2:nonconvex}
    \begin{align}
    \nonumber     & \text{maximize} \;\; \text{trace}(ZW)\\
    & \text{subject to}: G_{22}(k)\succ0 \qquad k\in\{0,1,...,N-1\}, \label{prob2:a}\\
    & G_{11}(0)-W-G_{12}(0)G_{22}^{-1}(0)G_{12}^T(0)\succeq 0, \label{prob2:b}\\
   \nonumber &E^T(k)\Big(G_{11}(k+1)-G_{12}(k+1)G_{22}^{-1}(k+1)G_{12}^T(k+1)\Big)\\
    &\times E(k)-G(k)+\Lambda(k)\succeq 0 \label{prob2:c}\\
    \nonumber & \forall k\in\{0,1,...,N-2\},\\
    &E^{T}(N-1)Q_{f}E(N-1)-G(N-1)+\Lambda(N-1)\succeq 0. \label{prob2:d}
    \end{align}
    \end{subequations}

    \end{prob}

\begin{proof}
First note that boundedness of the dual function \eqref{eq:19} from below requires that for \(k=0,1,...,N-2\),
\begin{equation}
    \Lambda(k)-G(k)+A_K^T(k)G(k+1)A_K(k)-H(k)=0,
    \label{eq:21}
\end{equation}
\begin{equation}
    \begin{split}
    &\Lambda(N-1)-G(N-1)+E^{T}(N -1)Q_{f}E(N-1)\\
    &-H(N-1)=0.
    \end{split}
    \label{eq:22}
\end{equation}
These conditions ensure that the Lagrangian $\mathcal{L}(K, S, G, H)$ is bounded below in the primal variables $S(k)$, since otherwise the trace terms would diverge to $-\infty$ as $S(k)$ becomes large. This is a standard necessary condition in duality theory to make the dual function finite \cite{boyd2004convex}.

Based on \eqref{eq:21} and \eqref{eq:22} and the fact that the \( H(k)\succeq0 \),
\begin{equation}
    \Lambda(k)+A_K^T(k)G(k+1)A_K(k)\succeq G(k),
    \label{eq:23}
\end{equation}
\begin{equation} \label{eq:24}
    \Lambda(N-1)+E^{T}(N-1)Q_{f}E(N-1)\succeq G(N-1),
\end{equation}
for \(k=0,1,...,N-2\). Let \( G(k)\) be partitioned as, 
\begin{equation}
    G(k)=\begin{bmatrix}
    G_{11}(k) \;& G_{12}(k)\\
    G_{12}^{T}(k) \;& G_{22}(k)
    \end{bmatrix}\in\mathbb{S}^{n+m},
    \label{eq:25}
\end{equation} 
where \( G_{11}(k)\in\mathbb{S}^{n} \), \( G_{12}(k)\in\mathbb{R}^{n\times m} \) and \( G_{22}(k)\in\mathbb{S}^{m} \). Based on this  partitioning, the last term of the Lagrangian \eqref{eq:18}, as the only nonzero term, can be rewritten as,
\begin{align} \label{eq:26}
    & \text{trace}\bigg(Z\begin{bmatrix}
    I_{n}\\
    K(0)
    \end{bmatrix}^{T}G(0)
    \begin{bmatrix}
    I_{n}\\
    K(0)
    \end{bmatrix}\bigg)\\
  \nonumber  & =\text{trace}\Big(Z\big(G_{11}(0)+G_{12}(0)K(0)+K^T(0)G^{T}_{12}(0)\\
  \nonumber  &\;\;\;\; +K^T(0)G_{22}(0)K(0)\big)\Big).
\end{align}
The expression in \eqref{eq:26} is bounded provided that,
\begin{equation}\label{eq:27}
    G_{22}(0)\succ0.
\end{equation}
This ensures the quadratic form in $K(0)$ is strictly convex, guaranteeing a unique minimizer exists. Accordingly, the the minimizer of \eqref{eq:26} would be characterized by initial feedback gain $ K^*(0)$ given by,
\begin{align} \label{eq:29}
    K^{*}(0)=-G^{-1}_{22}(0)G^{T}_{12}(0).
\end{align}
This expression results from setting the derivative of the quadratic term in \eqref{eq:26} with respect to $K(0)$ equal to zero.

Moreover, substituting,
\begin{equation}
    A_K(k)
    =\begin{bmatrix}
    I_{n}\\
    K(k+1)
    \end{bmatrix}E(k),
    \label{eq:30}
\end{equation}
into the constraint \eqref{eq:23} yields,
\begin{align}\label{eq:103}
    & E^T(k)\Big[G_{22}(k+1)K(k+1)+G_{12}^T(k+1)\Big]^T \\
 \nonumber   &\times G_{22}^{-1}(k+1) \Big[G_{22}(k+1)K(k+1)+G_{12}^T(k+1)\Big] E(k) \\
  \nonumber  &+E^T(k)\big(G_{11}(k+1)-G_{12}(k+1)G_{22}^{-1}(k+1)\\
 \nonumber   &\times G_{12}^T(k+1)\big)E(k)-G(k)+\Lambda(k)\succeq0.
\end{align}
By choosing,
\begin{equation}
    K^{*}(k)=-G^{-1}_{22}(k)G^{T}_{12}(k),
    \label{eq:31}
\end{equation}
for \(k=1,2,...,N-1\), the first positive semi-definite term of \eqref{eq:103} would be equal to zero and the constraint \eqref{eq:23} reduces to, 
\begin{align}\label{eq:33}
 \nonumber &  E^T(k)\big(G_{11}(k +1)-G_{12}(k+1)G_{22}^{-1}(k+1)\\
    &\times G_{12}^T(k+1)\big)E(k)-G(k)+\Lambda(k)\succeq0,
\end{align}
for \(k=0,1,...,N-2\). Placing \eqref{eq:29} in \eqref{eq:26}, the dual function is equivalent to,
\begin{equation}\label{eq:34}
    g(G,H)=\text{trace}\Big(Z\big(G_{11}(0)-G_{12}(0)G_{22}^{-1}(0)G_{12}^T(0)\big)\Big).
\end{equation}
Equation \eqref{eq:34} clearly defines the dual objective as a concave function in terms of $G_{11}$, $G_{12}$, and $G_{22}$.

Finally, we introduce the slack variable $W\in\mathbb{S}^{n\times n}$ such that,
\begin{equation}\label{eq:35}
    G_{11}(0)-G_{12}(0)G_{22}^{-1}(0)G_{12}^T(0)\succeq W.
\end{equation}
Recalling that $Z=\sum_{i=1}^n z_iz_i^T$ and the vectors $z_i$ are linearly independent, $Z$ is ensured to be positive-definite. Consequently, the objective function of the dual problem \eqref{eq:34} can be expressed as,
\begin{equation}\label{eq:36}
    \text{maximize~~}  \text{trace}(ZW),
\end{equation}
which is linear in $W$, with the constraints obtained in \eqref{eq:24}, \eqref{eq:27}, \eqref{eq:33} and \eqref{eq:35}. This completes the proof. \qed
\end{proof}

Accordingly, the dual problem \ref{problem2} can be equivalently reformulated as Problem \ref{problem3}, representing a convex optimization problem with a linear objective and LMI constraints, thus conforming to the standard framework of semidefinite programming.

\begin{prob}\label{problem3} Model-based SDP with variables  \( W\in\mathbb{S}^{n\times n} \) and \( G(k)\in\mathbb{S}^{n+m} \) indexed over \(k=0,1,...,N-1\).
\end{prob}

\begin{subequations}
    \begin{align} \label{eq:prob3}
 & \text{maximize~} trace(ZW)\\
   & \text{subject to}:
    \begin{bmatrix}
    G_{11}(0)-W\;\; & G_{12}(0)\\
    G^{T}_{12}(0) & G_{22}(0)
    \end{bmatrix}\succeq 0, \label{prob3:b}\\
   & \begin{bmatrix}
   \Delta(k) & E^{T}(k)G_{12}(k+1)\\
    G^{T}_{12}(k+1)E(k) & G_{22}(k+1)
    \end{bmatrix} \succeq 0, \label{prob3:c}\\
    \nonumber &\Delta(k)= E^{T}(k)G_{11}(k+1)E(k)-G(k)+\Lambda(k),\\
    \nonumber& \forall k \in \{0, 1, \dots, N-2\},\\
    &E^{T}(N-1)Q_{f}E(N-1)-G(N-1)+\Lambda(N-1)\succeq 0.
\end{align}
\end{subequations}

To show the equivalence between Problem~\ref{problem2} and Problem~\ref{problem3}, it is sufficient to apply the Schur complement~\cite{boyd2004convex} to constraint~\eqref{prob2:b} with respect to the block $G_{22}(0) \succ 0$, which yields~\eqref{prob3:b}. Similarly, applying the Schur complement to constraint~\eqref{prob2:c} with respect to the block $G_{22}(k+1) \succ 0$ for all $k \in \{0, 1, \dots, N-2\}$ results in~\eqref{prob3:c}. The subsequent theorem establishes that the optimal solution \( \{K^*(0),K^*(1),...,K^*(N-1)\}\) can be obtained by solving the convex problem \ref{problem3}.

\begin{theorem}
The optimal solutions of the semidefinite program introduced in Problem \ref{problem3}  are of the form of \eqref{eq:72} and \eqref{eq:74} Moreover, the optimal sequence of state-feedback control gains for \eqref{eq:100} is given by \eqref{eq:1011}, for all  $k=0,1,...,N-1$.
\end{theorem}

\begin{proof}

We prove optimality of the proposed solution by verifying that it satisfies the KKT conditions for Problem~\ref{problem2}. Given the inherent convexity of the dual problem, validating the satisfaction of the KKT conditions at a feasible point in Problem \ref{problem2} is adequate to establish optimality \cite{boyd2004convex}.

Consider the candidate solution given by,
\begin{equation}
\begin{split}
   & \hat{G}(k)= \left[\begin{array}{l}
           Q(k) + \gamma A^T(k)P(k+1)A(k) \\
         \gamma B^T(k)P(k+1)A(k)
    \end{array} \right.\\
        & ~~~~~~~~~~~~~~~~~~~\left.\begin{array}{l}
               \gamma A^T(k)P(k+1)B(k) \\
         R(k) + \gamma B^T(k)P(k+1)B(k)\\
        \end{array}\right],\\
\end{split}
\label{eq:72}
\end{equation}
\begin{equation}\label{eq:74}
     \hat{W}  =P(0),~~~~~~~~~~~~~~~~~~~~~~~~~~~~~~~~~~~~~~~~~~~~~~~~
\end{equation}
where $P(k)$ is the solution of the backward Riccati recursion \eqref{eq:73} with $P(N) = Q_f$.

Let $M_1(k),~M_2,~M_3(k)$ be the Lagrange multipliers associated with the three constraints in Problem 2. The KKT conditions corresponding to Problem~\ref{problem2} are detailed in Appendix~A and are represented by the constraints given in \eqref{eq:39}--\eqref{eq:45} and \eqref{eq:49}--\eqref{eq:55}. To assess the fulfillment of the KKT conditions, we examine the candidate Lagrange multiplier,
\begin{align*}
    & M_1(k)=0, ~~ k=0,1,...,N-1,\\
    & M_2= Z \succ 0,\\
    & M_3(k)= \begin{bmatrix}
    I_n\\
    -\hat{G}^{-1}_{22}(k)\hat{G}^T_{12}(k)
    \end{bmatrix}\Gamma(k)
    \begin{bmatrix}
    I_n\\
    -\hat{G}^{-1}_{22}(k)\hat{G}^T_{12}(k)
    \end{bmatrix}^T,\\
    & \Gamma(0)=Z,\\
    & \Gamma(k)=E(k-1)M_3(k-1)E^T(k-1).
\end{align*}
(i) Feasibility: It is straightforward to verify that the candidate solution $(\hat{G}(k), \hat{W})$ satisfies all primal constraints in Problem~\ref{problem2}.\\
(ii) Dual feasibility:  The constructed multipliers satisfy $M_1(k) \succeq 0$, $M_2 \succeq 0$, and $M_3(k) \succeq 0$ for all $k$, as required by \eqref{eq:39}--\eqref{eq:41}.\\
(iii) Complementary slackness: Condition \eqref{eq:42} is trivially satisfied by $M_1(k) = 0$. For \eqref{eq:43}, direct substitution shows,
\begin{align}\label{eq:755}
   & \hat{G}_{11}(0)-\hat{W}-\hat{G}_{12}(0)\hat{G}_{22}^{-1}(0)\hat{G}_{12}^T(0)\\
  \nonumber  &= Q(0)+A^T(0)P(1)A(0)-P(0)-A^T(0)P(1)B(0)\\
  \nonumber  &\times \big(R(0)+B^T(0)P(1)B(0)\big)^{-1}B^T(0)P(1)A(0)=0.
\end{align}

Similarly, \eqref{eq:44} holds because,
\begin{equation}
    \begin{split}
    &E^T(k)\big(\hat{G}_{11}(k+1)-\hat{G}_{12}(k+1)\hat{G}_{22}^{-1}(k+1)\hat{G}_{12}^T(k+1)\big)\\
    &\times E(k)-\hat{G}(k)+\Lambda(k)=E^T(k)P(k+1)E(k)-\hat{G}(k)\\
    &+\Lambda(k)=E^T(k)P(k+1)E(k)-E^T(k)P(k+1)E(k)\\
    &-\Lambda(k)+\Lambda(k)=0,
    \end{split}
    \label{eq:76}
\end{equation}
for \(k=1,2,...,N-2\). Finally, fulfillment of the complementary slackness condition \eqref{eq:45} follows from the fact that,
\begin{equation}
   E^T(N-1) Q_f E(N-1) - \hat{G}(N-1) + \Lambda(N-1) = 0.
\end{equation}
(iv) Stationarity: The KKT stationarity conditions \eqref{eq:49}--\eqref{eq:55} are satisfied by construction. Specifically, the derivative of the Lagrangian with respect to $W$ and each block of $G(k)$ for $k=0, 1, \ldots, N-1$ vanishes at the candidate solution. This follows from the structure of $\hat{G}(k)$ and $\hat{W}$, and the choice of multipliers $M_2$ and $M_3(k)$.\\
Accordingly, the candidate solution \( \hat{G}(k) ,\hat{W}\) in \eqref{eq:72} and \eqref{eq:74} satisfy all KKT conditions in Appendix A.\\

Furthermore, as the obtained solution remains unaffected by \(Z\), it can be substituted with any positive definite matrix, like the identity matrix. By the equivalence of Problem 2 and Problem 3, the optimal solutions of the SDP introduced in Problem \ref{problem3}  are of the form of \eqref{eq:72} and \eqref{eq:74}.\\
Accordingly, the time-varying controller $K^*(k)$ using the optimal solution derived from Problem 3, is given as,
\begin{align} \label{eq:1011}
 \nonumber   K^*(k)&= -\hat{G}_{22}^{-1}(k)\hat{G}_{12}^T(k)\\
 \nonumber & = -\big(R(k)+\gamma B^T(k)P(k+1)B(k)\big)^{-1} \\
    & ~~~~\times\gamma B^T(k)P(k+1)A(k),
\end{align}
which is consistent with the solution of the time-varying Riccati recursion for LQ problems. \qed.
\end{proof}

\begin{remark}
The optimal dual solution $\hat{G}(k)$ obtained in Theorem~2 corresponds to the $Q$-function matrix $\mathcal{Q}(k)$ representing the state-action value function at time $k$ for the finite-horizon LQ problem in the time-varying setting \cite{bertsekas2015dynamic}, i.e., $\mathcal{Q}(k) = \hat{G}(k), \quad \forall k \in \{0, 1, \dots, N-1\}$. Moreover, the optimal dual solution $\hat{W}$ gives the initial value function matrix, i.e., optimal cost-to-go matrix at time zero. To see this, let  $P(k)$ denote the state value function matrix at time $k$, so that the optimal cost-to-go from state $x(k)$ is $V(k) = x^T P(k) x(k)$. Given this, the Q-function at time $k$, defined as the cost of applying control  $u(k)$ in state $x(k)$ followed by the optimal policy thereafter, is,

\begin{align*}
    &Q_k(x(k), u(k)) =   x^T(k) Q(k) x(k)+ u^T(k) R(k) u(k)\\
   & + \gamma x^T(k+1) P(k+1) x(k+1).
\end{align*}

Substituting the dynamics into the terminal value term yields,
\begin{align*}
  & x^T(k+1) P(k+1) x(k+1) =\\
  & \begin{bmatrix} x(k) \\ u(k) \end{bmatrix}^T 
\begin{bmatrix} A^T(k) \\ B^T(k) \end{bmatrix}
P(k+1)
\begin{bmatrix} A(k) & B(k) \end{bmatrix}
\begin{bmatrix} x(k) \\ u(k) \end{bmatrix}.
\end{align*}
Combining all terms, the Q-function takes the quadratic form,
\begin{equation*}
  Q_k(x(k), u(k)) = \begin{bmatrix} x(k) \\ u(k) \end{bmatrix}^T \mathcal{Q}(k) \begin{bmatrix} x(k) \\ u(k) \end{bmatrix},  
\end{equation*}
with the matrix $\mathcal{Q}(k) \in \mathbb{S}_+^{n+m}$ given by,

\begin{equation}
\begin{split}
   & \mathcal{Q}(k)= \left[\begin{array}{l}
           Q(k) + \gamma A^T(k)P(k+1)A(k) \\
         \gamma B^T(k)P(k+1)A(k)
    \end{array} \right. \\
        & ~~~~~~~~~~~~~~~~~~~\left.\begin{array}{l}
               \gamma A^T(k)P(k+1)B(k) \\
         R(k) +  \gamma B^T(k)P(k+1)B(k)\\
        \end{array}\right].\\
\end{split}
\end{equation}

Comparing the state-action value matrix $\mathcal{Q}(k)$  with the KKT solution $\hat{G}(k)$ gives $\mathcal{Q}(k)=\hat{G}(k)$ for all $ k \in \{0, 1, \dots, N-1\}$. The optimal policy at each time step $k$ minimizes the Q-function with respect to the control input $u(k)$. Given the quadratic form of $Q_k(x(k), u(k))$, this minimizer can be found analytically. Partitioning $\mathcal{Q}(k)$ as, 
\begin{equation*}
    \mathcal{Q}(k) =
\begin{bmatrix}
\mathcal{Q}_{xx}(k) & \mathcal{Q}_{xu}(k) \\
\mathcal{Q}_{ux}(k) & \mathcal{Q}_{uu}(k)
\end{bmatrix},
\end{equation*}
the optimal control input is obtained as,
\begin{equation}
    \label{eq:optimal_gain}
   u^*(k) = -\mathcal{Q}_{uu}^{-1}(k) \mathcal{Q}_{ux}(k) \, x(k).
\end{equation}

This defines the optimal time-varying feedback gain as,

\begin{align*}
    & K^*(k) = -\big(R(k)+\gamma B^T(k)P(k+1)B(k)\big)^{-1}\\ & ~~~~~~~~~~~\times\gamma B^T(k)P(k+1)A(k).
\end{align*}

\end{remark}

\section{Data-ensemble-based approach for finite-horizon LQ control} \label{sec4}
In this section, the model-based SDP problem presented in Section \ref{sec3} is utilized to formulate an equivalent model-free optimization problem with LMI constraints. The resulting data-based LMI constraints are then employed to determine the optimal set of control gains, all accomplished without necessitating access to the time-varying system dynamics.\\

\subsection{Data-driven reformulation of finite-horizon LQ Control for LTV systems}

Let \( x_{j}(k)\) and \( u_{j}(k)\) represent the data collected during an experiment \( j \) at time \( k \), where \(j=1,2,...,l\). Consider,
\begin{equation}
    X_k=\gamma^{\frac{k}{2}}\begin{bmatrix}
    x_1(k) \;& x_2(k) \;& ... \;& x_l(k)
    \end{bmatrix},
    \label{eq:80}
\end{equation}
\begin{equation}
    U_k=\gamma^{\frac{k}{2}}\begin{bmatrix}
    u_1(k) \;& u_2(k) \;& ... \;& u_l(k)
    \end{bmatrix},
    \label{eq:81}
\end{equation}
and,
\begin{equation}
    D_k=\begin{bmatrix}
    X_k\\
    U_k
    \end{bmatrix}\in\mathbb{R}^{(n+m)\times l},
    \label{eq:82}
\end{equation}
for \(k=0,1,...,N-1\).
According to the principles of matrix congruence, as outlined in \cite{linearalgebra}, if the matrix $D_k$ is full row rank, i.e., $\text{rank}(D_k)=m+n$, then it is valid to pre- and post-multiply both sides of the inequality in \eqref{eq:33} by \(D_k^T\) and \(D_k\), respectively, for all \(k = 0, 1, \dots, N-2\). This operation yields the equivalent constraint,
\begin{align*}
    &D^T_kE^T(k)\big(G_{11}(k+1)-G_{12}(k+1)G_{22}^{-1}(k+1)\\
     &G_{12}^T(k+1)\big)E(k)D_k-D^T_k\big(G(k)-\Lambda(k)\big)D_k\succeq0.
\end{align*}
Note that $ E(k)D_k=\sqrt{\gamma}\begin{bmatrix}
    A(k) \;& B(k)
    \end{bmatrix}D_k=X_{k+1}$. Thus, the last constraint can be equivalently represented as,
\begin{align} \label{eq:84}
 \nonumber   & X^T_{k+1}\big(G_{11}(k+1)-G_{12}(k+1)G_{22}^{-1}(k+1)\\
    &G_{12}^T(k+1)\big)X_{k+1}-D^T_k\big(G(k)-\Lambda(k)\big)D_k\succeq0.
\end{align}
Likewise, the application of left and right multiplication by \( D^T_{N-1} \) and \( D_{N-1} \)  to inequality \eqref{eq:24} results in,
\begin{align*}
     &D^T_{N-1}E^{T}(N-1)Q_{f}E(N-1)D_{N-1}\\
    &-D^T_{N-1}\big(G(N-1)-\Lambda(N-1)\big)D_{N-1}\succeq 0,
\end{align*}
which is equivalent to,
\begin{equation}    \label{eq:85}
    X^T_{N}Q_{f}X_N-D^T_{N-1}\big(G(N-1)-\Lambda(N-1)\big)D_{N-1} \succeq 0.
\end{equation}

The equivalence between \eqref{eq:84} and \eqref{eq:85} and their model-based counterparts requires that   $D_k$ matrices for all $k=0,1,...,N-1$, must exhibit a full row rank. Each matrix $D_k$ aggregates input-state pairs from multiple independent experiments at the same time index. This condition ensures that the collected data is sufficiently rich to span the input-state space and captures the behavior of the system at that time. Unlike the LTI case, where a single global data matrix suffices to capture the full system behavior \cite{de2019formulas}, the LTV case necessitates a collection of such local conditions across the horizon.

\begin{remark}
Ensuring the full-rank condition $\mathrm{rank}(D_k) = n + m$  for all $k \in \{0, \dots, N-1\}$ in practice requires careful design of input signals and experimental protocols. Two key requirements must be satisfied: 
(i) the number of independent experiments $l$ used to construct $D_k$ must be at least $n + m$, and 
(ii) the input-state samples $\begin{bmatrix} x_j(k) \\ u_j(k) \end{bmatrix}$, collected across these experiments $j = 1, \dots, l$, must be linearly independent. To meet these conditions, it is advisable to initialize each experiment from a distinct initial state and apply persistently exciting input sequences of sufficiently high order to adequately stimulate all system modes. Excitation signals such as Gaussian white noise, sums of sinusoids, and
and frequently switching piecewise constant inputs are commonly employed to ensure sufficient spectral richness \cite{JIANG20122699}. Furthermore, it is important to diversify the inputs across experiments to effectively span the local input-state space at each time step. These practices are consistent with the classical notion of persistent excitation in system identification and control
~\cite{willems2005note}. While the full theoretical extension of the Fundamental Lemma to the LTV setting remains an active area of research, the modular rank-based framework proposed here serves as a practical approach to data-driven control of time-varying systems.
\end{remark}  
By restating the inequality \eqref{eq:84} in the LMI form, we reach the condition,
\begin{equation}
    \begin{split}
    &\begin{bmatrix}
    \Delta(k) \;& X^T_{k+1}G_{12}(k+1)\\
    G^{T}_{12}(k+1)X_{k+1} \;& G_{22}(k+1)
    \end{bmatrix}\succeq 0,\\
  & \Delta(k)=X^T_{k+1}G_{11}(k+1)X_{k+1}-D^{T}_k\big(G(k)-\Lambda(k)\big)D_k,
    \end{split}
    \label{eq:86}
\end{equation}
for \(k=0,1,...,N-2\). Finally, the model-free SDP problem is described as Problem 4.

\begin{prob}\label{problem4} Model-free SDP problem with variables  \( W\in\mathbb{S}^{n\times n} \) and \( G(k)\in\mathbb{S}^{n+m} \) indexed over \(k=0,1,...,N-1\).
\end{prob}
\begin{align} \label{eq:prob4}
& \text{maximize} \;\; trace(W)\\
\nonumber & \text{subject  to}:
\begin{bmatrix}
G_{11}(0)-W\;\; & G_{12}(0)\\
G^{T}_{12}(0) & G_{22}(0)
\end{bmatrix}\succeq 0,\\
\nonumber &\begin{bmatrix}
    \Delta(k) \;& X^T_{k+1}G_{12}(k+1)\\
    G^{T}_{12}(k+1)X_{k+1} \;& G_{22}(k+1)
    \end{bmatrix}\succeq 0,\\
\nonumber  & \Delta(k)=X^T_{k+1}G_{11}(k+1)X_{k+1}-D^{T}_k\big(G(k)-\Lambda(k)\big)D_k,\\
\nonumber& \forall k \in \{0, 1, \dots, N-2\},\\
\nonumber &X^T_{N}Q_{f}X_N-D^T_{N-1}\big(G(N-1)-\Lambda(N-1)\big)D_{N-1}\succeq 0.
\end{align}

The optimization problem formulated in \eqref{eq:prob4} is entirely derived by data matrices. It requires \( l \geq n+m \) batches of input-state data sequence, each of length \( N \). To minimize the number of required batches, we choose \( l=n+m \) and construct \( X_k \) and \( D_k \) matrices from the data batches according to \eqref{eq:80}-\eqref{eq:82}. 

\begin{remark}
Consider the special case where the dynamics of the LTV system, as expressed by equation \eqref{eq:1}, adheres to a T-periodic pattern,
\begin{equation*}
    A(k+T)=A(k),~~~B(k+T)=B(k),~~~ \forall k \geq 0.
\end{equation*}
In this context, the data richness condition for an ensemble of $l = n + m$  input-state trajectories can be satisfied by a single sufficiently long experiment that encompasses at least $n + m$ periods of the system. This allows the repeated structure of the system dynamics to be exploited, thereby enabling the use of one extended trajectory to extract the required data matrices at each relevant time instant.
\end{remark}

Although periodicity can simplify data collection by reducing the number of required experiments, the proposed algorithm is not restricted to periodic systems and is applicable to general LTV systems, provided that sufficiently informative data from multiple independent trajectories are available through physical experiments or simulations. Interestingly, this requirement closely parallels standard assumptions in the system identification of LTV systems, where identification is typically carried out using a subspace framework based on an ensemble of input-state data \cite{identification}. In fact, LTV identification using multi-experiment data commonly treats each time step $k$ as a separate regression problem. Theoretical results in the literature confirm that, if for each $k$, the matrix formed by stacking input-state data from all experiments at that time instant has full row rank, then the time-varying system matrices  $A_k$ and $B_k$ can be uniquely identified. These rank conditions serve as natural generalizations of the persistency of excitation condition from the LTI setting to the time-varying case.

\subsection{Special Case: Finite-horizon LQ control of LTI systems}
The proposed framework is applicable to finite-time LQ control of LTI systems, as well. In this specific scenario, it can be demonstrated that a single trajectory data batch, with a length of $m+n+1$, is adequate for constructing data-driven LMIs and generating the time-dependent control gain matrix $K(k)$. Given Problem \ref{problem2}, it is straightforward to show that the model-based equivalent optimization problem for the LTI system with a finite-horizon time-invariant performance index simplifies to,
\begin{equation}\label{eq:LTIcase1}
    \begin{split}
    & \text{maximize} \;\; \text{trace}(W)\\
    & \text{subject to}: G_{22}(k)\succ0,\\
    & G_{11}(0)-W-G_{12}(0)G_{22}^{-1}(0)G_{12}^T(0)\succeq 0,\\
    &E^T\Big(G_{11}(k+1)-G_{12}(k+1)G_{22}^{-1}(k+1)G_{12}^T(k+1)\Big)\\
    &\times E-G(k)+\Lambda\succeq0 \qquad \forall k\in\{0,1,...,N-2\},\\
    &E^{T}Q_{f}E-G(N-1)+\Lambda\succeq 0.
    \end{split}
\end{equation}
with, $ \Lambda= \begin{bmatrix}
    Q \;& 0\\
    0 \;& R
    \end{bmatrix}$, and $E=\sqrt\gamma\begin{bmatrix}
    A \;\;& B
    \end{bmatrix}$.
Assume that a single trajectory of length $s$ is to be collected with the initial sample indexed by $k_0$. Construct the data matrix $L \in \mathbb{R}^{(n+m)\times s}$ as follows,
\begin{equation} \label{Ldef}
L:=\begin{bmatrix}
x(k_0) & x(k_0+1) & \cdots & x(k_0+s-1) \\
u(k_0) & u(k_0+1) & \cdots & u(k_0+s-1)
\end{bmatrix}.
\end{equation}
Define $X_L \in \mathbb{R}^{n\times s}$ as,
\begin{equation}\label{Xdef}
\begin{aligned}
X_L:=&\begin{bmatrix} A \quad B \end{bmatrix}L\\
=&\left[\begin{matrix}
Ax(k_0)+Bu(k_0) \mkern6mu Ax(k_0+1)+Bu(k_0+1) \end{matrix}\right.\\
&\qquad \left.\begin{matrix} \cdots \mkern6mu Ax(k_0+s-1)+Bu(k_0+s-1)
\end{matrix}\right]\\
=&\begin{bmatrix}
x(k_0+1) & x(k_0+2) & \cdots & x(k_0+s)
\end{bmatrix}.
\end{aligned}
\end{equation}
If the column rank of matrix $L$ is $m+n$, it is possible to multiply the third inequality constraint of \eqref{eq:LTIcase1},
from left and right by $L^T$ and $L$, to obtain the equivalent set of model-free LMIs with the single trajectory data matrices $X_L$ and $L$, given as Problem 5. Ensuring a column rank of $m+n$ for matrix $L$ can be achieved by collecting $s+1=m+n+1$ samples from a single trajectory. Moreover, it is crucial for the input signal within the interval $[k_0, k_0+s-1]$ to demonstrate persistent excitation (PE) of a minimum order of $n+1$.

\begin{prob}\label{problem5} Model-free SDP for finite horizon LQ control of LTI systems with variables  \( W\in\mathbb{S}^{n\times n} \) and \( G(k)\in\mathbb{S}^{n+m} \) indexed over \(k=0,1,...,N-1\).
\end{prob}
\begin{align} \label{eq:prob5}
& \text{maximize} \;\; trace(W)\\
\nonumber & \text{subject to}:
\begin{bmatrix}
G_{11}(0)-W\;\; & G_{12}(0)\\
G^{T}_{12}(0) & G_{22}(0)
\end{bmatrix}\succeq 0,\\
\nonumber &\begin{bmatrix}
    \Delta(k) \;& \sqrt{\gamma} X^T_LG_{12}(k+1)\\
    \sqrt{\gamma} G^{T}_{12}(k+1)X_L \;& G_{22}(k+1)
    \end{bmatrix}\succeq 0,\\
\nonumber  & \Delta(k)=\gamma X^T_LG_{11}(k+1)X_L-L^{T}\big(G(k)-\Lambda\big)L,\\
\nonumber& \forall k \in \{0, 1, \dots, N-2\},\\
\nonumber &\gamma X^T_LQ_{f}X_L-L^T\big(G(N-1)-\Lambda\big)L\succeq 0.
\end{align}
The optimization problem presented in \eqref{eq:prob5} is completely independent of any reliance on a model. It solely necessitates a singular trajectory for input-state data, with a length of $m+n+1$.\\

\subsection{Comparison of sample efficiency and computational complexity with existing methods}

\begin{table*}[t]
\centering
\caption{Comparison of data-driven finite-horizon value function approximation methods.}
\label{tab:comparison1}
\footnotesize
\renewcommand{\arraystretch}{1}
\begin{tabular}{|
    >{\centering\arraybackslash}m{4.2cm}|
    >{\centering\arraybackslash}m{3.2cm}|
    >{\centering\arraybackslash}m{3.4cm}|
    >{\centering\arraybackslash}m{2.7cm}|
}
\hline
\textbf{Method} & \textbf{No. of Experiments} & \textbf{Rank Condition on Data Matrix} & \textbf{No. of Steps} \\
\hline
Off-policy policy iteration \cite{pang2018data, pang2019adaptive} & Sufficiently large (no analytical lower bound) & $\frac{m(m+1)}{2} + mn + \frac{n(n+1)}{2}$ & Iterative until convergence \\
\hline
Q-learning for LTV systems \cite{possieri2021iterative} & $\dfrac{(n+m)(n+m+1)}{2}$ & $n+m$ & $N$ backward steps \\
\hline
Q-learning for finite-horizon LTT systems \cite{calafiore2020output} & $\dfrac{(n+m)(n+m+1)}{2}$ & $n+m$ & $N$ backward steps \\
\hline
Proposed method (LTV, Problem 4) & $n+m$ & $n+m$ & Single step \\
\hline
Proposed method (LTI, Problem 5) & $1$ & $n+m$  & Single step \\
\hline
\end{tabular}
\end{table*}

\begin{table*}[t]
\centering
\caption{Comparison of data-driven methods: problem size and data requirements.}
\label{tab:comparison2}
\footnotesize
\renewcommand{\arraystretch}{1}
\begin{tabular}{|
    >{\centering\arraybackslash}m{3.8cm}|
    >{\centering\arraybackslash}m{1.8cm}|
    >{\centering\arraybackslash}m{4.6cm}|
    >{\centering\arraybackslash}m{1.2cm}|
    >{\centering\arraybackslash}m{1.6cm}|
}
\hline
\textbf{Method} & \textbf{No. of Samples} & \textbf{No. of Decision Variables} & \textbf{No. of LMIs} & \textbf{No. of Equality Constraints} \\
\hline
Data-driven LQ for LTV (covariance-based) \cite{nortmann2020data,nortmann2021direct} & $(n+m)(N+1)$ & $(n + m)nN + \frac{m(m + 1)}{2}N + \frac{n(n + 1)}{2}N$ & $2N+1$ & $n^2$ \\
\hline
Data-driven LQ for LTI (covariance-based) \cite{rotulo2020data} & $(n+m)(N+1)$ & $(n + m)nN + \frac{m(m + 1)}{2}N + \frac{n(n + 1)}{2}N$ & $2N+1$ & $n^2$ \\
\hline
Proposed method (LTV, Problem 4) & $(n + m)(N+1)$ & $\frac{(n + m)(n + m + 1)}{2}N + \frac{n(n + 1)}{2}$ & $N$  & 0 \\
\hline
Proposed method (LTI, Problem 5) & $n+m+1$ & $\frac{(n + m)(n + m + 1)}{2}N + \frac{n(n + 1)}{2}$ & $N$  & 0 \\
\hline
\end{tabular}
\end{table*}

To assess the data efficiency of the proposed method, we perform a comparative analysis against the off-policy policy iteration approach described in \cite{pang2018data, pang2019adaptive}, as well as the Q-learning method \cite{calafiore2020output} and  \cite{possieri2021iterative} for LQ state feedback control of finite horizon time-invariant and time-varying systems, respectively. 
Table~\ref{tab:comparison1} summarizes a comparative analysis of data-driven methods for finite-horizon value function approximation in discrete-time systems. The methods are evaluated in terms of the number of experiments required, the rank condition imposed on the corresponding data matrix, and the number of computational steps needed to evaluate the value functions. The off-policy policy iteration method proposed in \cite{pang2018data, pang2019adaptive}
requires the existence of a total number of experiments, denoted as $l_0$ for data collection such that for all $l> l_0$, the rank condition associated with the data matrix is  $ \frac{m(m+1)}{2}+mn+\frac{n(n+1)}{2}.$ In this method, the precise total number of required experiments $l$, or the confirmation of the existence of such $l$ for ensuring the data-dependent matrix rank condition, is not known a priori. Moreover, the approach proceeds through iterative updates until convergence. The Q-learning approaches for LTV and LTT systems proposed in \cite{possieri2021iterative} and \cite{calafiore2020output}, respectively, require $\frac{(n+m)(n+m+1)}{2}$ experiments along with solving $N$ backward semi-definite programs. This backward structure makes the algorithm more susceptible to error propagation, especially problematic for systems with long time horizons. In contrast, the proposed methods for both LTV systems (Problem~4) and LTT systems (Problem~5) significantly reduce the data and computational burden. For the LTV case we require only $n+m$ experiments to satisfy the rank condition and compute the value function in a single step, without recursive updates. Moreover, for the LTI case, a single rollout of length $n+m+1$ samples are sufficient to derive the globally optimal solution. This highlights the improved sample efficiency and reduced computational complexity offered by the proposed framework.

The proposed method is also compared with the data-driven finite-horizon LQ design proposed in \cite{nortmann2020data,nortmann2021direct} for the LTV systems and \cite{rotulo2020data} for the LTI case. Table~ \ref{tab:comparison2}  presents a comparison of problem sizes and data requirements among available data-driven finite-horizon optimal control methods. Note that the methods in \cite{nortmann2020data,nortmann2021direct, rotulo2020data} are all based on convariance selection problem without any interpretation from the perspective of state-action value functions (Q-farctors).  The methods are evaluated based on the number of samples required, the total number of decision variables, the number of LMIs, and the number of equality constraints.

The proposed method for both LTV and LTI systems demonstrates a substantial reduction in computational complexity compared to existing covariance-based approaches  \cite{nortmann2020data, nortmann2021direct, rotulo2020data}. In particular, the number of decision variables in our method is given by
$\frac{(n + m)(n + m + 1)}{2}N + \frac{n(n + 1)}{2}$. In contrast, the method in \cite{nortmann2021direct} introduces a significantly larger number of decision variables: $(n + m)nN + \frac{m(m + 1)}{2}N + \frac{n(n + 1)}{2}N$.
Assuming  $n>m$, the reduction in decision variables is quantified by $\Delta = n^2 N - \frac{n(n + 1)}{2},$
which is strictly positive for all $N > 1 $, emphasizing the superior efficiency of the proposed approach. This advantage becomes more pronounced for larger state dimensions or longer horizons.

In terms of constraints, our method requires only $N$ LMIs and imposes no equality constraints, while the covariance-based methods require $2N+1$ LMIs and $n^2$ equality constraints.

A particularly notable advantage arises in the LTI case, where our method (as a special case of LTV systems in Problem 5) requires only a single rollout of length $n + m + 1$ to collect all the necessary data for optimization. This stands in sharp contrast to existing LTI methods presented in \cite{rotulo2020data}—a special case of the more general LTV frameworks discussed in \cite{nortmann2020data, nortmann2021direct}—which typically require a significantly larger number of data samples. In this context, the proposed formulation is particularly well-suited for finite-horizon LTI scenarios where data acquisition is costly or constrained.

\section{Simulation results} \label{sec5}
In this section, the performances of the proposed data-driven approach is evaluated and compared with the recent $Q$-function approximation method for LTV systems \cite{possieri2021iterative}. The first example examines an unstable linear time-varying system, while the second example focuses on stabilization of an unstable nonlinear system by utilizing merely the state and input trajectories.

\begin{example}
Consider the unstable LTV system,
\makeatletter
\newcommand{\vast}{\bBigg@{5}}
\newcommand{\Vast}{\bBigg@{6}}
\makeatother
\begin{small}
\begin{equation}
    \begin{split}
    &A(k)=\vast[\begin{matrix}
    1-0.05k \;\;& -0.5cos(0.2k)\sqrt{k} \\
    -0.1cos(0.3k) \;\;& 0 \\
    0.1sin(0.5k) \;\;& -0.2 \\
    0 \;\;& 0.01k
    \end{matrix}\\
    &\begin{matrix}
    -0.1 \;\;& 0.02k \\
    0.03\sqrt{k} \;\;& 0.2 \\
    1-0.01k \;\;& 0.002k \\
    0.1cos(0.2k) \;\;& 1+0.03sin(0.1k)
    \end{matrix}\vast],\\
    &B(k)=0.1\begin{bmatrix}
    k \;\;& 1-cos(0.1k) \;\;& k \;\;& 1
    \end{bmatrix}^T.
    \end{split}
    \label{eq:88}
\end{equation}
\end{small}
\end{example}

\begin{figure}[h]
    \centering
    \begin{subfigure}[b]{8.4cm}
        \centering
        \includegraphics[width=8.4cm]{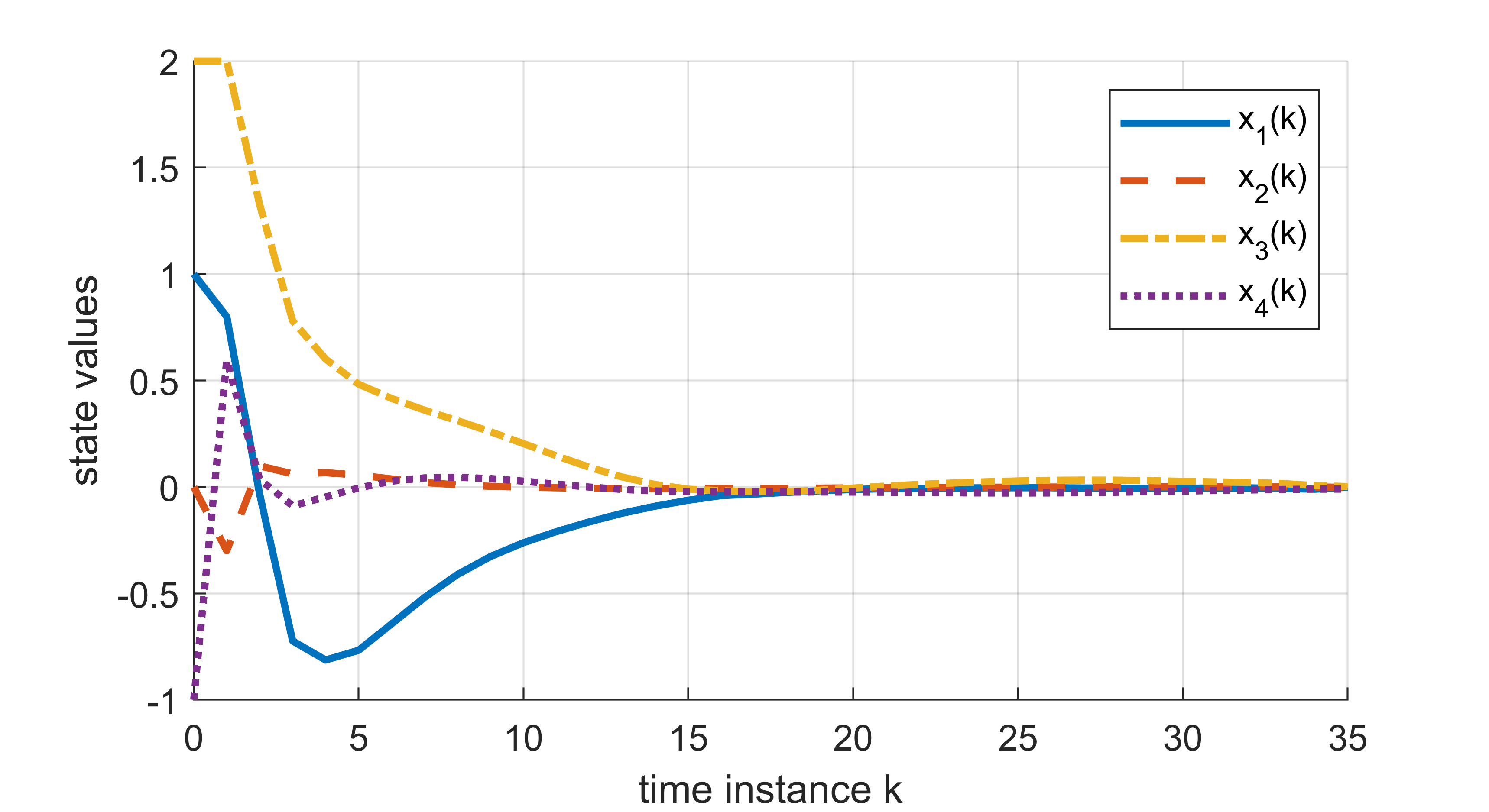}
        \caption{}
        \label{fig:state_mf}
    \end{subfigure}
    \vspace{0.5em}
    \begin{subfigure}[b]{8.4cm}
        \centering
        \includegraphics[width=8.4cm]{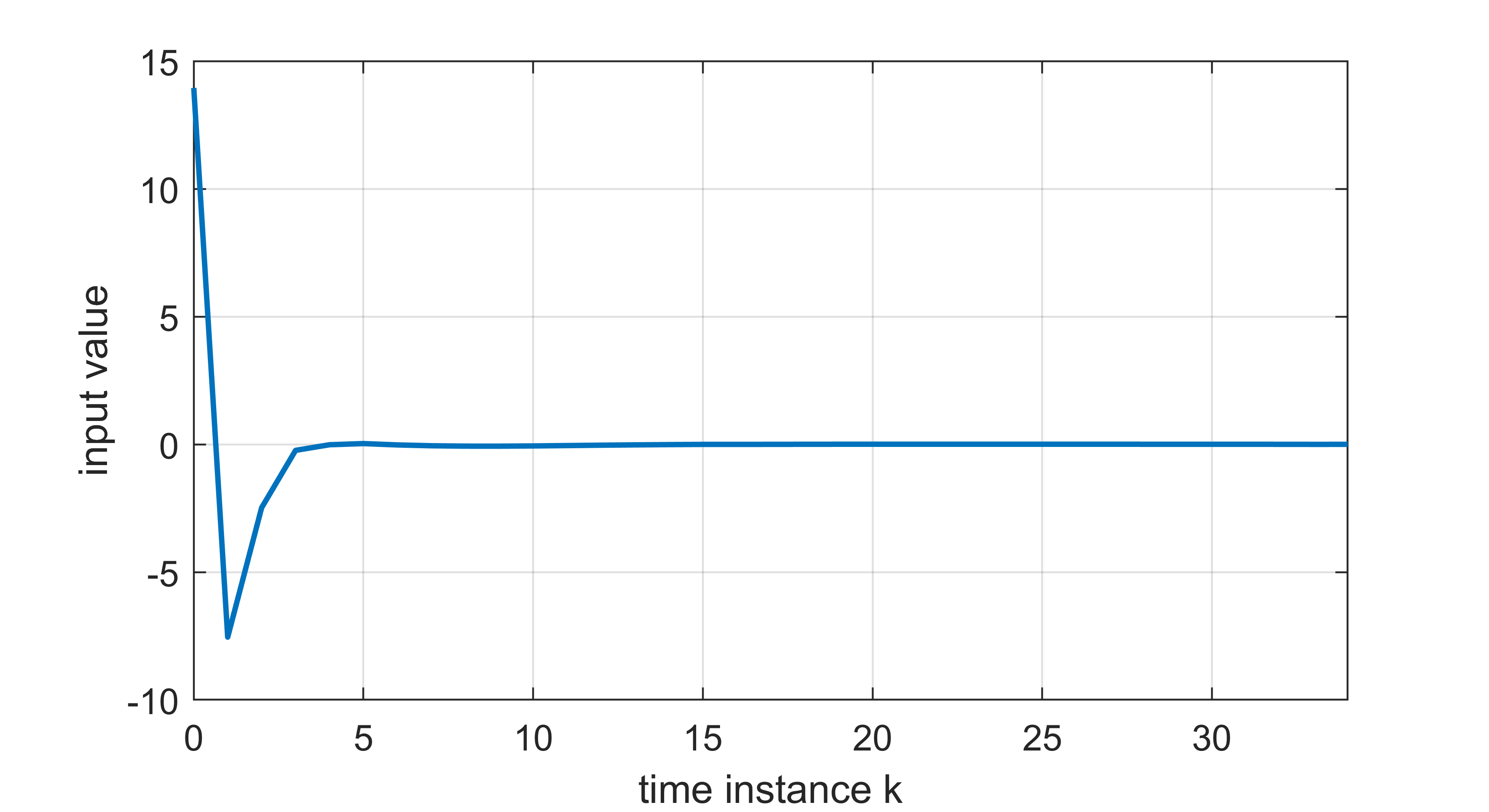}
        \caption{}
        \label{fig:input_mf}
    \end{subfigure}
    \caption{
    (a) Closed-loop state trajectories and (b) control signal for Example 1 using the proposed model-free method (Problem 4) with noise-free data.
    }
    \label{fig:mf_example}
\end{figure}

\noindent\textbf{A. Verification of the global optimality of the proposed approach:} Model-based and model-free LQ designs are implemented for the system described in Problem~3 and Problem~4, respectively, with parameters set to \( N = 35 \) and \( \gamma = 0.98 \). The performance weighting matrices are chosen as \( Q(k) = I_4 \), \( R(k) = 0.01 \), and \( Q_f = 10I_4 \). The state trajectories and control input generated by the proposed model-free method are shown in Figures~\ref{fig:state_mf} and~\ref{fig:input_mf}, with the initial state $x_0 = \begin{bmatrix} 1 & 0 & 2 & -1 \end{bmatrix}^T$, respectively. Remarkably, the control gains and resulting trajectories from the proposed model-free approach match exactly those obtained using the model-based time-varying Riccati recursion with the objective value $J(x_0)= 19.4674$. As a result, the model-based trajectories are not redundantly displayed. This exact correspondence demonstrates that the proposed method attains globally optimal performance in this setting.

\noindent\textbf{B. Comparison with the $Q$-learning method under noise-free data:}
We compare the performance of the proposed single-shot method (Problem~4) against the iterative $Q$-function approximation method for LTV systems introduced in~\cite{possieri2021iterative}. To ensure a fair and consistent evaluation, we consider a noise-free scenario and set the discount factor $\gamma = 1$ in our method to match the undiscounted cost formulation used in~\cite{possieri2021iterative}.

As shown in Table~\ref{tab:noExp}, our method consistently achieves a near-optimal LQ cost (around 20.1175) across all ensemble sizes, reflecting the ability to solve the finite-horizon problem globally in a single step. In contrast, the method of~\cite{possieri2021iterative} exhibits highly suboptimal performance for small to moderate numbers of input-state trajectories and shows acceptable performance as the number of experiments increases. Notably, even as the number of experiments increases to 30, the method fails to reach the global optimum; its best cost plateaus at 34.5284, remaining significantly above the optimal value. This demonstrates that the backward recursive $Q$-function approximation in~\cite{possieri2021iterative} may not guarantee convergence to the global solution, even under ideal, noise-free conditions.

\begin{figure}[h]
\includegraphics[width=8.4cm]{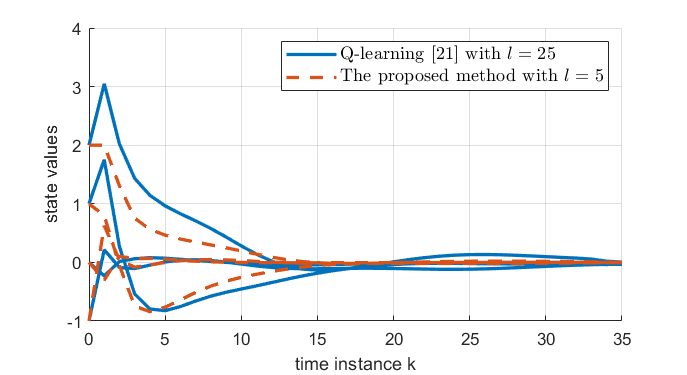}
\captionof{figure}{Comparison between the state trajectories of the closed-loop system by the proposed method in Problem 4 ($l=5$) and the method in~\cite{possieri2021iterative} ($l=25$) with noise-free data.}
\label{fig:compare}
\end{figure}

Figure~\ref{fig:compare} presents the state trajectories of the controlled system using the proposed method with $l=5$, in comparison to the $Q$-function approximation approach from~\cite{possieri2021iterative}, evaluated at its best performance with $l=25$. The results demonstrate the superior performance of the proposed method, which achieves near-optimal control with significantly lower ensemble size, owing to its guaranteed global optimality.

\begin{table}[h!]
    \centering
    \caption{Comparison of LQ performance between the proposed single-shot method (Problem~4) and the $Q$-learning method with backward SDPs in~\cite{possieri2021iterative}, across different ensemble sizes ($l$).}
    \label{tab:noExp}
    \renewcommand{\arraystretch}{1.2}
    \setlength{\tabcolsep}{8pt}
    \begin{tabular}{|c|c|c|
    } 
    \hline
    $l$ & $J(x_0)$ (Prob.~4) & $J(x_0)$ (\cite{possieri2021iterative})  \\
    \hline
    5 & 20.1157 & 464.5485 \\
    10 & 20.1157 & $1.2546\times10^3$ \\
    15 & 20.1161 & $3.1033\times10^3$ \\
    20 & 20.1166 & 47.2599 \\
    25 & 20.1174 & 34.5284 \\
    30 & 20.1191 & 34.5284 \\
    \hline
    \end{tabular}
\end{table}

\noindent\textbf{C. Comparison with the $Q$-learning method under noisy data:} To assess robustness under measurement noise, Table~\ref{tab:noise} reports the average LQ cost and computation time, each obtained over 100 independent Monte Carlo simulations under reported levels of measurement noise. The noise is additive and drawn from a Gaussian distribution  $\mathcal{N}(0, \sigma^2)$, where $\sigma$ denotes the standard deviation. The proposed method uses $l=5$ experiments, while the $Q$-learning-based method of~\cite{possieri2021iterative} uses $l=25$, consistent with their original settings. The results indicate a significant degradation in performance for the method of~\cite{possieri2021iterative}, particularly as noise levels increase. This is primarily due to its higher sensitivity to measurement noise and the compounding effect of error propagation in the sequential backward optimization process, where earlier decisions rely on the accuracy of solutions computed for later time steps.

\begin{figure}[h]
\includegraphics[width=8.4cm]{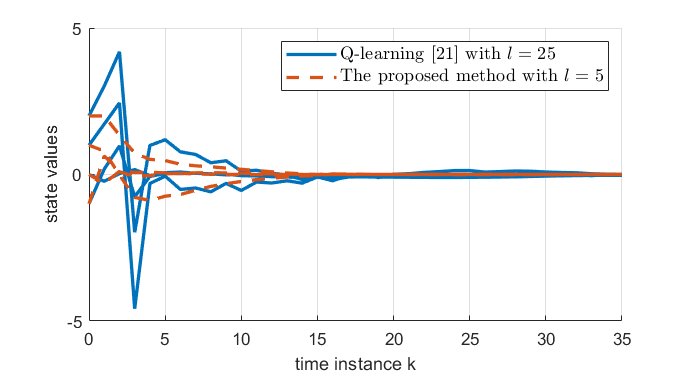}
\captionof{figure}{Comparison of state trajectories of the closed-loop system using the proposed method in Problem~4 ($l = 5$) and the method from~\cite{possieri2021iterative} ($l = 25$), with additive measurement noise $\sigma = 0.001$ applied to the data.}
\label{fig:noise}
\end{figure}

Figure~\ref{fig:noise} compares the state trajectories obtained using the proposed method ($l=5$, $\sigma = 0.001$) with those from the approach in~\cite{possieri2021iterative} ($l=25$, $\sigma = 0.001$). Despite relying on a smaller ensemble size, our method exhibits greater robustness to noise and achieves a significantly lower LQ cost ($J(x_0) = 32.87$) compared to~\cite{possieri2021iterative}($J(x_0) = 87.23$). Furthermore, the proposed method requires substantially less computation time: only $1.96$ seconds versus $15.32$ seconds for ~\cite{possieri2021iterative}.

\begin{table}[h!]
\centering
\caption{Average LQ cost and computation time over 100 Monte Carlo simulations, comparing the proposed method (Problem~4) with the $Q$-learning approach from~\cite{possieri2021iterative}, under varying levels of Gaussian noise.}
\label{tab:noise}
\resizebox{\columnwidth}{!}{
\begin{tabular}{|c|c|c|c|c|
}
\hline
$\sigma$ & $J(x_0)$ & $J(x_0)$ & Time (s) & Time (s) \\
         & Prob.~4  & \cite{possieri2021iterative} & Prob.~4  & \cite{possieri2021iterative} \\
\hline
0       & 20.1157 & 34.5284 & 1.96 & 15.32 \\
0.0005  & 20.1354 & 34.7410 & 3.12 & 24.05 \\
0.001   & 32.8797 & 87.2345 & 2.80 & 29.33 \\
0.005   & 59.8128 & $3.13\times10^3$ & 1.82 & 13.04 \\
0.01    & 129.0405 & $4.12\times10^3$ & 1.89 & 14.67 \\
0.02    & 179.6365 & $5.99\times10^6$ & 1.71 & 20.82 \\
\hline
\end{tabular}
}
\end{table}

\begin{example}
The LTV framework's ability in representing nonlinear systems with surrogate models makes the results in this paper a step forward in facilitating data-driven control of nonlinear systems, as well. Accordingly, in this example, we explore the prospect of achieving stabilization of a closed-loop system for time-varying, open-loop unstable nonlinear dynamics while adhering to the desired LQ performance, leveraging the developed data-dependent LMIs. The nonlinear dynamics under consideration are represented as,

\begin{equation}
    \begin{split}
    &x_1(k+1)=-\big(1+0.04cos(0.1k)\big)x_1(k)\\
    &+ 0.5sin(0.2k)u(k)\\
    &x_2(k+1)=-0.1sin(0.5k)x_1(k)x_2(k)\\
    &- 0.05k\frac{x_3(k)}{\big(x_2(k) + 1\big)}-0.05u(k)\\
    &x_3(k+1)=0.04k^{\frac{3}{2}}x_1(k)-x_2(k)\\
    &+0.1kx_2(k)u(k)
    \end{split}
    \label{eq:89}
\end{equation}
\begin{figure}[h]
\includegraphics[width=8.4cm]{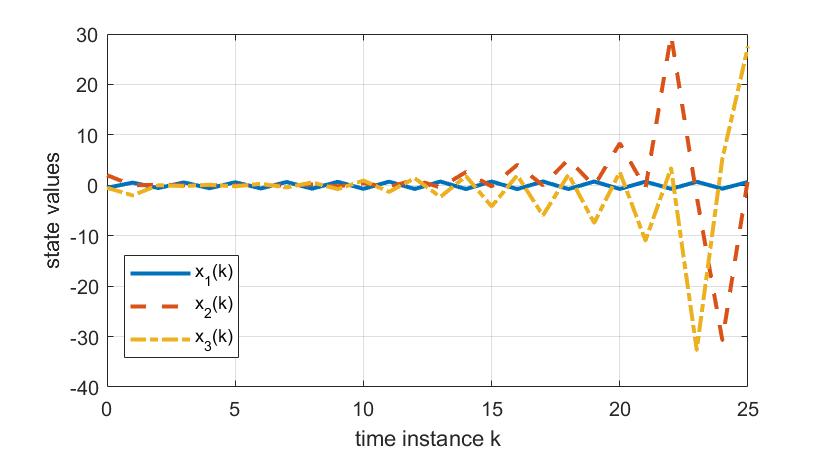}
\captionof{figure}{State trajectories of the non-linear open loop system in Example 2.}
\label{fig:nonlin_open}
\end{figure}

\begin{figure}[h]
    \centering
    \begin{subfigure}[b]{8.4cm}
        \centering
        \includegraphics[width=8.4cm]{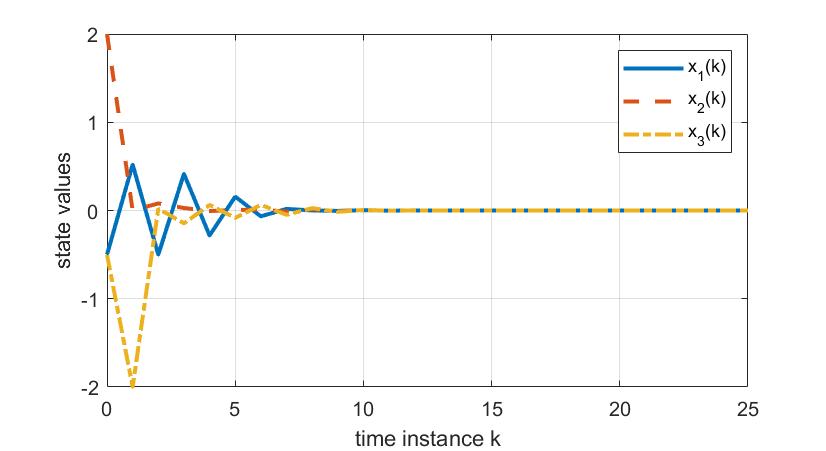}
        \caption{}
        \label{fig:state_nonlin}
    \end{subfigure}
    \vspace{0.5em}
    \begin{subfigure}[b]{8.4cm}
        \centering
        \includegraphics[width=8.4cm]{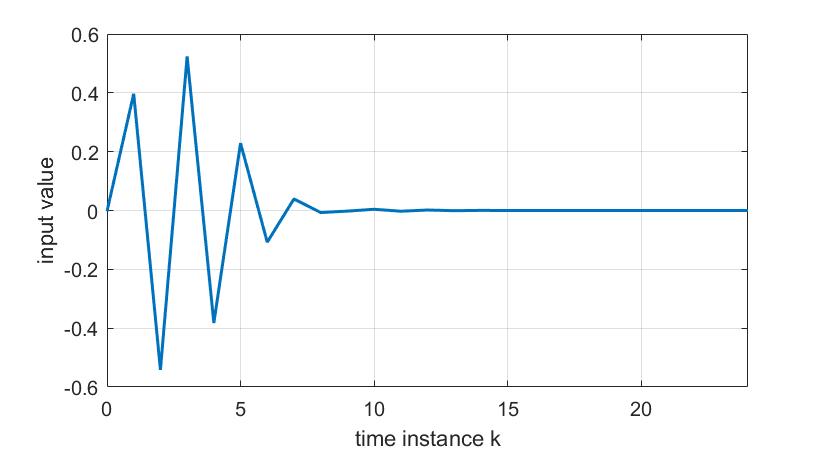}
        \caption{}
        \label{fig:input_nonlin}
    \end{subfigure}
    \caption{
     (a) Closed-loop state trajectories and (b) control signal for nonlinear system in Example 2 using the proposed model-free method (Problem 4).
    }
    \label{fig:nonlin_example}
\end{figure}
The state trajectories of the open-loop system are shown in Figure \ref{fig:nonlin_open}. Since the proposed data-driven formulation is completely independent of the linear state-space matrices, it provides opportunities to stabilize nonlinear systems whose trajectories remain close enough to an unstable equilibrium point. In this example, we set \( N=25\), \( Q(k)=(0.5k+2)I_4\), \( R(k)=2-0.01k\), \( Q_f=50I_4\) and \( \gamma=0.98\). The state and input trajectories are gathered and the data-dependent LMIs in Problem 4 are solved. The simulation results starting from initial condition $[-0.5,2,-0.5]$ with the obtained control signal sequence are reported in Figures \ref{fig:state_nonlin} and \ref{fig:input_nonlin}. Note that in this example, linearization followed by a model-based and time-varying Riccati equation design fails to provide a stabilizing solution. These results confirm the effectiveness of the proposed data-driven approach in stabilizing nonlinear systems with satisfactory performance.
\end{example}

\section{Conclusion} \label{sec6}
 This study introduces a new data-driven approach to address the finite horizon LQ problem in LTV systems. The proposed methodology entails the non-iterative solution of data-dependent semi-definite programs, 
 by an ensemble of input-state trajectories
 from a minimum of $m+n$ experiments. Additionally, it is shown that, in the case of LTI systems, the direct data-driven finite-horizon problem can be solved using a single trajectory of length $m+n+1$. Simulation results on LTV systems demonstrate clear superiority over LTV Q-function approximation in terms of both optimality and sample efficiency. Additional simulations on nonlinear systems suggest that the method may also aid stabilization in more complex settings. This is enabled by the LTV framework’s ability to approximate nonlinear dynamics with surrogate models, marking a step toward data-driven control of nonlinear systems using linear constructs.\\
While the proposed framework ensures global optimality under nominal conditions, it does not explicitly guarantee robustness to measurement noise. Extending it to robust settings is a promising direction, given its natural alignment with LMI-based robust control methods. Another key avenue is developing an informativity-based framework for LTV systems, analogous to the LTI theory in~\cite{van2020data}, by incorporating time-local data conditions and time-varying dynamics. Moreover, as LTV models can approximate a wide range of nonlinear systems via local linearization or surrogate modeling, the approach also lays groundwork for sample-efficient, data-driven control of nonlinear systems.

\section*{Appendix A: The KKT condition for Problem 2} \label{sec7}

The Lagrangian function corresponding to Problem 2 is,

\begin{align} \label{eq:38}
  \nonumber      &\mathcal{L}(G,W,M_1,M_2,M_3)=-\text{trace}(ZW)\\
      \nonumber    &-\sum_{k=0}^{N-1}\text{trace}\big(M_1(k)G_{22}(k)\big)\\
      \nonumber    &-\text{trace}\Big(M_2\big(G_{11}(0)-W-G_{12}(0)G_{22}^{-1}(0)G_{12}^T(0)\big)\Big)\\
     \nonumber     &-\sum_{k=0}^{N-2}\text{trace}\bigg(M_3(k)\Big(E^T(k)\big(G_{11}(k+1)-G_{12}(k+1)\\
     \nonumber     &\times G_{22}^{-1}(k+1)G_{12}^T(k+1)\big)E(k)-G(k)+\Lambda(k)\Big)\bigg)\\
      \nonumber    &-\text{trace}\Big(M_3(N-1)\big(E^{T}(N-1)Q_{f}E(N-1)\\
        &-G(N-1)+\Lambda(N-1)\big)\Big).
\end{align}
To derive the KKT conditions, first note that according to dual feasibility constraints, we have,
\begin{align}
    &  M_1(k)\succeq0~~ k=0,1,...,N-1, \label{eq:39}\\
    & M_2\succeq0, \label{eq:40}\\
    &  M_3(k)\succeq0 ~~ k=0,1,...,N-1,  \label{eq:41}
\end{align}
Next, the Complementary slackness requires that,
\begin{equation}
  \text{trace}\left(  M_1(k){G}_{22}(k)\right)=0,
    \label{eq:42}
\end{equation}
\begin{equation}
  \text{trace}\left(  M_2\Big[{G}_{11}(0)-{W}-{G}_{12}(0){G}_{22}^{-1}(0){G}_{12}^T(0)\Big]\right)=0,
    \label{eq:43}
\end{equation}
\begin{align}\label{eq:44}
\nonumber & \text{trace}\left(  M_3(k)\Big[E^T(k )\big({G}_{11}(k+1)-{G}_{12}(k+1){G}_{22}^{-1}(k+1)\right.\\
   & \left.\times {G}_{12}^T(k+1)\big)E(k)-{G}(k)+\Lambda(k)\Big]\right)=0,
\end{align}
\begin{align}\label{eq:45}
\nonumber & \text{trace}\left(  M_3(N-1)\Big[E^{T}(N-1)Q_{f}E(N  -1)-{G}(N-1)\right.\\
    &\left. +\Lambda(N-1)\Big]\right)=0.
    \end{align}

Since the complementary slackness condition \eqref{eq:42} yields $ M_1(k) = 0 $ for all $ k \in \{0, 1, \dots, N-1\} $, the associated terms vanish and are thus omitted in the expressions for the stationarity conditions.

Next, the Stationarity conditions require that,
\begin{equation*} 
\nabla_W\mathcal{L}=0, \quad
\nabla_{G(0)}\mathcal{L}=0, \quad \nabla_{G(k)}\mathcal{L}=0.
\end{equation*}
To interpret these stationarity conditions, we can rewrite ${G}(k)$ for \(k=0,1,...,N-1\) as, \begin{equation*}
    \begin{split}
    {G}(k)=&T_1{G}_{11}(k)T^T_1+T_1{G}_{22}(k)T^T_2+T_2{G}^T_{12}(k)T^T_1\\
    &+T_2{G}_{22}(k)T^T_2,
    \end{split}
\end{equation*}
with $ T_1=\begin{bmatrix}
    I_{n}\\
    0
    \end{bmatrix}\in\mathbb{R}^{(n+m)\times n},\;\;
    T_2=\begin{bmatrix}
    0\\
    I_{m}
    \end{bmatrix}\in\mathbb{R}^{(n+m)\times m}
    \label{eq:47}$.
Thus, the stationarity conditions are described as, \eqref{eq:49}-\eqref{eq:55}:
\begin{equation}
    \nabla_W\mathcal{L}=-Z+M_2=0,~~~~~~~~~~~~~~~~~~~~~~~~~~~~~~~~~~~~~
    \label{eq:49}
\end{equation}
\begin{equation}
    \nabla_{G_{11}(0)}\mathcal{L}=-M_2
    +T^T_1M_3(0)T_1=0,~~~~~~~~~~~~~~~~~~~~~
    \label{eq:50}
\end{equation}
\begin{align}\label{eq:51}
  &  \nabla_{G_{12}(0)}\mathcal{L}=M_2{G}_{12}(0)\hat{G}^{-1}_{22}(0)
    +T^T_1M_3(0)T_2=0,
\end{align}
\begin{equation}
    \begin{split}
    \nabla_{G_{22}(0)}\mathcal{L}=&-{G}^{-1}_{22}(0){G}^T_{12}(0)M_2{G}_{12}(0){G}^{-1}_{22}(0)~~~~~~~~~~\\
    &+T^T_2M_3(0)T_2=0,
    \end{split}
    \label{eq:52}
\end{equation}
and for \(k=1,2,...,N-1\),
\begin{equation}
    \begin{split}
    \nabla_{G_{11}(k)}\mathcal{L}=&-E(k-1)M_3(k-1)E^T(k-1)~~~~~~~~~~~~~~~~~~~~\\
    &+T^T_1M_3(k)T_1=0,
    \end{split}
    \label{eq:53}
\end{equation}
\begin{equation}
    \begin{split}
    \nabla_{G_{12}(k)}\mathcal{L}=&E(k-1)M_3(k-1)E^T(k-1)\\
    &\times {G}_{12}(k){G}^{-1}_{22}(k)+T^T_1M_3(k)T_2=0,~~~~~~~~~~~
    \end{split}
    \label{eq:54}
\end{equation}
\begin{equation}
    \begin{split}
    \nabla_{G_{22}(k)}\mathcal{L}&=-{G}^{-1}_{22}(k)\hat{G}^T_{12}(k)E(k-1)M_3(k-1)\\
    &\times E^T(k-1){G}_{12}(k){G}^{-1}_{22}(k)+T^T_2M_3(k)T_2=0.
    \end{split}
    \label{eq:55}
\end{equation}
\bibliographystyle{elsarticle-num}
\bibliography{mybibliography}

\begin{thebibliography}{10}
\expandafter\ifx\csname url\endcsname\relax
  \def\url#1{\texttt{#1}}\fi
\expandafter\ifx\csname urlprefix\endcsname\relax\def\urlprefix{URL }\fi
\expandafter\ifx\csname href\endcsname\relax
  \def\href#1#2{#2} \def\path#1{#1}\fi

\bibitem{markovsky2023data}
I.~Markovsky, L.~Huang, F.~D{\"o}rfler, Data-driven control based on the behavioral approach: From theory to applications in power systems, IEEE Control Systems Magazine 43~(5) (2023) 28--68.

\bibitem{van2020data}
H.~J. Van~Waarde, J.~Eising, H.~L. Trentelman, M.~K. Camlibel, Data informativity: a new perspective on data-driven analysis and control, IEEE Transactions on Automatic Control 65~(11) (2020) 4753--4768.

\bibitem{kiumarsi2014reinforcement}
B.~Kiumarsi, F.~L. Lewis, H.~Modares, A.~Karimpour, M.-B. Naghibi-Sistani, Reinforcement {Q}-learning for optimal tracking control of linear discrete-time systems with unknown dynamics, Automatica 50~(4) (2014) 1167--1175.

\bibitem{li2018off}
X.~Li, Z.~Peng, L.~Liang, Off-policy {Q}-learning for infinite horizon {LQR} problem with unknown dynamics, in: 2018 IEEE 27th international symposium on industrial electronics (ISIE), IEEE, 2018, pp. 258--263.

\bibitem{lee2018primal}
D.~Lee, J.~Hu, Primal-dual {Q}-learning framework for {LQR} design, IEEE Transactions on Automatic Control 64~(9) (2018) 3756--3763.

\bibitem{li2022model}
M.~Li, J.~Qin, W.~X. Zheng, Y.~Wang, Y.~Kang, Model-free design of stochastic {LQR} controller from a primal--dual optimization perspective, Automatica 140 (2022) 110253.

\bibitem{fazel2018global}
M.~Fazel, R.~Ge, S.~Kakade, M.~Mesbahi, Global convergence of policy gradient methods for the linear quadratic regulator, in: International Conference on Machine Learning, PMLR, 2018, pp. 1467--1476.

\bibitem{hu2022towards}
B.~Hu, K.~Zhang, N.~Li, M.~Mesbahi, M.~Fazel, T.~Ba{\c{s}}ar, Towards a theoretical foundation of policy optimization for learning control policies, arXiv preprint arXiv:2210.04810 (2022).

\bibitem{de2019formulas}
C.~De~Persis, P.~Tesi, Formulas for data-driven control: Stabilization, optimality, and robustness, IEEE Transactions on Automatic Control 65~(3) (2019) 909--924.

\bibitem{van2021matrix}
H.~J. van Waarde, M.~K. Camlibel, A matrix finsler’s lemma with applications to data-driven control, in: 2021 60th IEEE Conference on Decision and Control (CDC), IEEE, 2021, pp. 5777--5782.

\bibitem{farjadnasab2022model}
M.~Farjadnasab, M.~Babazadeh, Model-free {LQR} design by {Q}-function learning, Automatica 137 (2022) 110060.

\bibitem{de2021low}
C.~De~Persis, P.~Tesi, Low-complexity learning of linear quadratic regulators from noisy data, Automatica 128 (2021) 109548.

\bibitem{calafiore2020output}
G.~C. Calafiore, C.~Possieri, Output feedback {Q}-learning for linear-quadratic discrete-time finite-horizon control problems, IEEE transactions on neural networks and learning systems 32~(7) (2020) 3274--3281.

\bibitem{Liu2017BatchtoBatchFL}
S.-J. Liu, M.~Krsti{\'c}, T.~Başar, Batch-to-batch finite-horizon lq control for unknown discrete-time linear systems via stochastic extremum seeking, IEEE Transactions on Automatic Control 62 (2017) 4116--4123.

\bibitem{scheinker2021extremum}
A.~Scheinker, D.~Scheinker, Extremum seeking for optimal control problems with unknown time-varying systems and unknown objective functions, International Journal of Adaptive Control and Signal Processing 35~(7) (2021) 1143--1161.

\bibitem{yang2018efficient}
Y.~Yang, An efficient {LQR} design for discrete-time linear periodic system based on a novel lifting method, Automatica 87 (2018) 383--388.

\bibitem{pang2020reinforcement}
B.~Pang, Z.-P. Jiang, I.~Mareels, Reinforcement learning for adaptive optimal control of continuous-time linear periodic systems, Automatica 118 (2020) 109035.

\bibitem{qu2021stable}
G.~Qu, Y.~Shi, S.~Lale, A.~Anandkumar, A.~Wierman, Stable online control of linear time-varying systems, in: Learning for Dynamics and Control, PMLR, 2021, pp. 742--753.

\bibitem{pang2018data}
B.~Pang, T.~Bian, Z.-P. Jiang, Data-driven finite-horizon optimal control for linear time-varying discrete-time systems, in: 2018 IEEE Conference on Decision and Control (CDC), IEEE, 2018, pp. 861--866.

\bibitem{pang2019adaptive}
B.~Pang, T.~Bian, Z.-P. Jiang, Adaptive dynamic programming for finite-horizon optimal control of linear time-varying discrete-time systems, Control theory and technology 17~(1) (2019) 73--84.

\bibitem{possieri2021iterative}
C.~Possieri, G.~P. Incremona, G.~C. Calafiore, A.~Ferrara, An iterative data-driven linear quadratic method to solve nonlinear discrete-time tracking problems, IEEE Transactions on Automatic Control 66~(11) (2021) 5514--5521.

\bibitem{fong2018dual}
J.~Fong, Y.~Tan, V.~Crocher, D.~Oetomo, I.~Mareels, Dual-loop iterative optimal control for the finite horizon {LQR} problem with unknown dynamics, Systems $\&$ Control Letters 111 (2018) 49--57.

\bibitem{nortmann2020data}
B.~Nortmann, T.~Mylvaganam, Data-driven control of linear time-varying systems, in: 2020 59th IEEE Conference on Decision and Control (CDC), IEEE, 2020, pp. 3939--3944.

\bibitem{nortmann2021direct}
B.~Nortmann, T.~Mylvaganam, Direct data-driven control of linear time-varying systems, IEEE Transactions on Automatic Control 68~(8) (2023) 4888--4895.

\bibitem{rotulo2020data}
M.~Rotulo, C.~De~Persis, P.~Tesi, Data-driven linear quadratic regulation via semidefinite programming, IFAC-PapersOnLine 53~(2) (2020) 3995--4000.

\bibitem{willems2005note}
J.~C. Willems, P.~Rapisarda, I.~Markovsky, B.~L. De~Moor, A note on persistency of excitation, Systems $\&$ Control Letters 54~(4) (2005) 325--329.

\bibitem{verhoek2021fundamental}
C.~Verhoek, R.~T{\'o}th, S.~Haesaert, A.~Koch, Fundamental lemma for data-driven analysis of linear parameter-varying systems, in: 2021 60th IEEE Conference on Decision and Control (CDC), IEEE, 2021, pp. 5040--5046.

\bibitem{bertsekas2015dynamic}
D.~P. Bertsekas, Dynamic programming and optimal control 4th edition, volume ii, Athena Scientific (2015).

\bibitem{kirk2004optimal}
D.~E. Kirk, Optimal control theory: an introduction, Courier Corporation, 2004.

\bibitem{boyd2004convex}
S.~P. Boyd, L.~Vandenberghe, Convex optimization, Cambridge university press, 2004.

\bibitem{linearalgebra}
G.~Strang, Linear Algebra and Its Applications, 1968.

\bibitem{JIANG20122699}
Y.~Jiang, Z.-P. Jiang, Computational adaptive optimal control for continuous-time linear systems with completely unknown dynamics, Automatica 48~(10) (2012) 2699--2704.

\bibitem{identification}
M.~Verhaegen, X.~Yu, A class of subspace model identification algorithms to identify periodically and arbitrarily time-varying systems, Automatica 31~(2) (1995) 201--216.

\end{thebibliography}
\end{document}